\documentclass{emulateapj}

\shortauthors{TOFFLEMIRE, BURKHART \& LAZARIAN}
\shorttitle{TSALLIS STATISTICS OF ISM TURBULENCE}
\submitted{ApJ Submitted}
\begin{document}

\title{Interstellar Sonic and Alfv\'enic Mach Numbers and the Tsallis Distribution}

\author{Benjamin M. Tofflemire\altaffilmark{1}, 
Blakesley Burkhart\altaffilmark{2}, 
A. Lazarian\altaffilmark{2}}

\altaffiltext{1}{Astronomy Department, University of Washington, Box 351580
  Seattle, WA 98195}
\altaffiltext{2}{ Astronomy Department, University of Wisconsin, Madison, 475
  N. Charter St., WI 53711, USA}

\begin{abstract}
In an effort to characterize the Mach numbers of ISM magnetohydrodynamic (MHD) 
turbulence,  we study the probability distribution functions (PDFs) of spatial 
increments of density, velocity, and magnetic field for fourteen ideal 
isothermal MHD simulations at resolution $512^3$. In particular, we fit the PDFs using the Tsallis 
function and study the dependency of fit parameters on the compressibility and 
magnetization of the gas. We find that the Tsallis function fits PDFs of MHD 
turbulence well, with fit parameters showing
sensitivities to the sonic and Alfv\'en Mach numbers. For 3D
density, column density, and position-position-velocity (PPV) data we 
find that the amplitude and width of the PDFs shows a dependency on the 
sonic Mach number. We also find the width of the PDF is sensitive 
to global Alfv\'enic Mach number especially in cases where the sonic number is high. These 
dependencies are also found for mock observational cases, where cloud-like 
boundary conditions, smoothing, and noise are introduced. The ability of Tsallis 
statistics to characterize sonic and Alfv\'enic Mach numbers of simulated ISM turbulence 
point to it being a useful tool in the analysis of the 
observed ISM, especially when used simultaneously with other statistical techniques.
\end{abstract}

\section{INTRODUCTION}
\label{intro}
Understanding the dynamics and evolution of the interstellar medium (ISM) is
critical to advancing our knowledge of many astrophysical phenomena spanning 
a wide range of scales such as star
formation, cosmic ray physics, magnetic reconnection, galaxy evolution, and
magnetic dynamo theory (Elmegreen \& Scalo 2004). An essential component of the current paradigm of the
ISM is the ubiquitous existence of magnetohydrodynamic (MHD) turbulence (see
review by McKee \& Ostriker 2007). Turbulence and magnetic fields play a
crucial roll in each of these processes and are some of the main drivers of
ISM evolution.  Evidence for the role of turbulence is seen in
the ``big power law'' of the electron density fluctuations (Armstrong, Rickett, \& Spangler 1995), 
fractal structure in the molecular media (Elmegreen \& Falgarone 1996, Stutzki et al. 1998),
and intensity fluctuations contributed by both density and turbulent velocity
in channel maps (Crovisier \& Dickey 1983; Green 1993; 
Deshpande, Dwarakanath, \& Goss 2000; Elmegreen, Kim, \& Staveley-Smith 2001).

Despite the obvious importance of MHD turbulence to astrophysics, few methods
exist to study it directly. Due to advances in computational power and the
general recognition of turbulence as an important ISM process,
 major advances have been made in MHD
turbulence theory and observation in the last ten years. However, the issue
of turbulence and its effects on the ISM (and processes therein) remains one of the most
exciting and open problems in the field (Elmegreen \& Scalo 2004).

Astrophysical turbulence is a complex nonlinear phenomena
that can occur in a multiphase media with many energy injection
sources on scales ranging from kpc down to sub-AU.
Although limited in complexity, numerical simulations of turbulence provide one of the best avenues
for researchers of the ISM to understand the nature of magnetized
turbulence. The combined efforts of predictive theory and numerical tests have
greatly increased our knowledge of MHD turbulence, including its
 anisotropy, intermittency, and imbalanced nature
 (see Cho, Lazarian, \& Vishniac 2002, Kowal \& Lazarian 2010, Beresnyak \& Lazarian 2010).

Yet what about observationally driven studies of turbulence? The most common
observational techniques to study turbulence include scintillation studies,
which are limited to fluctuations in only ionized media (e.g.
 Narayan \& Goodman 1989; Spangler \& Gwinn 1990), density fluctuations
via column density maps, and radio spectroscopic observations via centroids of
spectral lines (Falgarone et al. 1994; Miesch \& Bally 1994; Miesch \& Scalo
1995; Lis et al. 1998; Miesch, Scalo, \& Bally 1999). Column densities are the most abundant
and easily obtained observable data and have shown that density fluctuations can
be a useful and straightforward way of gauging turbulence parameters
(Monin \& Yaglom 1967; Lithwick \& Goldreich 2001; Cho \& Lazarian 2003).
Position-Position-Velocity (PPV) spectroscopic data has the
advantage over column density maps in that it contains information on the turbulent
velocity field. However, this type of data provides contributions of \textit{both}
density and velocity fluctuations entangled together, and the process of
separating the two has proven a challenging problem (see Lazarian 2006b).
In addition, one must use caution when dealing with PPV data, as structures seen
in velocity slices are not one-to-one with structures in three dimensional position space.

One of the main approaches for characterizing ISM turbulence is based on using
statistical techniques and descriptions. The most common ``go-to'' tool for both
observers and theorist alike is the spatial power spectrum. In fact, most of
the attempts to relate observations to models has been by obtaining the spectral
index (i.e. the log-log slope of the power spectrum) of column density and
velocity. While obtaining the spectral index of column density is
straightforward, more sophisticated techniques for obtaining the velocity
spectral index from PPV data have been recently developed.  These include: Velocity
Channel Analysis (VCA) (Lazarian \& Pogosyan 2000; Esquivel et al. 2003;
Lazarian \& Pogosyan 2004; Lazarian et al. 2001; Padoan et al. 2003; Chepurnov
\& Lazarian 2009), the Spectral Correlation Function (SCF) (Rosolowsky et
al. 1999; Padoan, Rosolowsky, \& Goodman 2001), Velocity Coordinate Spectrum (VCS)(Lazarian \&
Pogosyan 2008, 2006; Chepurnov \& Lazarian 2006, 2009;
Padoan et al. 2009), and Modified Velocity Centroids (Lazarian \& Esquivel 2003;
Esquivel \& Lazarian 2005; Ossenkopf et al. 2006; Esquivel et al. 2007). For
cases where turbulence is supersonic, the VCA is most appropriate while
centroids can be used in subsonic cases.

Although the power spectrum is useful for obtaining information about energy
transfer over scales, it does not provide a full picture of turbulence,
partially because it only contains information on Fourier amplitudes. An
example of this is illuminated in a study by Chepurnov et al. (2008), who showed
that a  substantially different distribution of density could have the same
power spectrum.  In light of this, many other techniques have been developed
to study and parametrize observational magnetic turbulence. These include
higher order spectrum, such as the bispectrum, higher order statistical
moments, topological techniques (such as genus), clump and hierarchical
structure algorithms (such as dendrograms), principle component analysis, and structure/correlation
functions as tests of intermittency and anisotropy (for examples of such
studies see Heyer \& Schloerb 1997, Burkhart et al. 2009; Chepurnov \& Lazarian 2009; 
Kowal, Lazarian, \&  Beresnyak 2007;
Goodman et al. 2009; Burkhart et al. 2011).  Wavelets methods, and variations on them such
as the $\Delta$-variance method, have also been shown to be very useful in characterizing
inhomogeneities in data (see Ossenkopf, Krips, \& Stutzki 2008a, 2008b).  

In particular, many of the
studies mentioned above focus on obtaining the parameters of turbulence from
observations. These parameters include sonic and Alfv\'enic Mach numbers,
injection scale, gas temperature, and Reynolds number.  In particular the sonic
and Alfv\'enic Mach numbers provide much coveted information on the gas
compressibility and magnetization.  Many of these techniques, geared towards
obtaining the parameters of turbulence via density fluctuations studies, were
successfully applied to observational data (see Burkhart et al. 2010 and
Chepurnov et al. 2008, for examples). VCA and VCS were also applied to
Galactic HI data and successfully recovered the spectrum of velocity (see
Chepurnov et al. 2010).

In this vein, Esquivel \& Lazarian (2010),  henceforth known as EL10, used the so-called
Tsallis statistic for studies of MHD turbulence. It is this
statistic that is the focus of this work, and here we will further illuminate
its uses. The Tsallis distribution is a  function that can be fit to incremental
PDFs of turbulent density, magnetic field, and velocity.
In astrophysical settings, Tsallis statistics was originally used in
the context of solar wind observations (Burlaga \& Vi\~nas 2004a). EL10 applied this method to
3D MHD simulations with four varying values of sonic and Alfv\'enic Mach
number at $256^3$ resolution. They explored density, magnetic field, velocity,
and column densities, and found that the Tsallis  distribution is a very good fit to PDFs of
increments of turbulence.  They also found that the parameters of the fit that
describe the width and tails of the PDFs showed dependency on the
compressibility and magnetization of the simulation.  The statistic is
particularly useful in that it is scale independent
and thus a comparison between the analysis of simulations and observations is not
burdened with complicated scaling relations. This opens up the possibility of
using the Tsallis method on observed ISM data in order to gain access to information
on these parameters. While EL10 was the first to implement this tool on simulations of ISM MHD turbulence,
they used low spatial resolution simulations,
a small parameter regime, and did not explore the dependencies on the amplitude fit parameter.
They also did not explore the use of Tsallis statistics
on spectroscopic data. In this paper, we will greatly extend their parameter
range and resolution from 4 at $256^3$ isothermal MHD simulations to 14 at
$512^3$.  In addition, we will more explicitly explore the ability of the Tsallis function
to describe data of an observable nature, such as smoothed synthetic column density maps
and synthetic PPV data of varying velocity resolution. We also investigate the
quality of our fits and subsequent fit parameters.

The paper is organized as follows.  In \S~\ref{tsallis} we describe the
Tsallis distribution which is fit to increments of MHD turbulence
PDFs. In \S~\ref{numericalsetup} we discuss our numerical scheme and
resulting simulations.  In \S~\ref{3d} we apply Tsallis to
non-observational 3D quantities (density and directional components of
magnetic field and  velocity) and test the accuracy of our fits. In \S~\ref{observ}
we apply this tool to observational quantities such as column density and PPV data. In
\S~\ref{disc} we discuss our results followed by the conclusions in
\S~\ref{conc}. 

\section{TSALLIS STATISTICS}
\label{tsallis}
The Tsallis distribution (Tsallis 1988) was originally derived as
a means to extend traditional Boltzmann-Gibbs mechanics to fractal and
multifractal systems. The complex dynamics of multifractal systems apply to
many natural environments such as ISM turbulence
(Shivamoggi 1995). It is therefore fitting to
explore the extent to which Tsallis statistics can be used to describe these
systems and processes therein (for further discussion see section \ref{relate}). 
Work of this nature was first carried out by Burlaga
and collaborators (Burlaga \& Vi\~{n}as 2004a, 2004b, 2005a, 2005b, 2006; 
Burlaga, Ness, \& Acu{\~n}as 2006, 2007, 2009; Burlaga, Vi{\~n}as, \& Wang 2007) 
to describe the temporal variation
in PDFs of magnetic field strength and velocity of solar wind measured by the
\emph{Voyager 1} \& \emph{2} spacecrafts. EL10 used Tsallis statistics to
describe the spatial variation in PDFs of density, velocity, and magnetic field of
 MHD simulations similar to those used here. Both efforts
found that the Tsallis distribution provided adequate fits to their PDFs 
and gave insight into statistics of turbulence.
The Tsallis function of an arbitrary incremental PDF $(\Delta f)$ has the form:

\begin{equation}
\label{(1)}
R_{q}= a \left[1+(q-1) \frac{\Delta f(r)^2}{w^2} \right]^{-1/(q-1)}
\end{equation}
The fit is described by the three dependent parameters $a$, $q$, and $w$. The
$a$ parameter describes the amplitude while $w$ is related to the width
or dispersion of the distribution. Parameter $q$, referred to as the ``non-extensivity
parameter'' or ``entropic index'', describes the sharpness and tail size of the
distribution.  

The arbitrary function used to describe density and the directional
components of velocity and magnetic field in this application takes the form of 
our incremental PDF. It has the form 
$\Delta f(r)=(f(r)-\langle f(r)\rangle_{\bf{x}})/\sigma_{f}$, where
$\langle$...$ \rangle_{\bf{x}}$
refers to a spatial average. Depending on the quantity in question, we set 
$f(x)=\rho(x + r) - \rho(x); 
v_{x}(x + r) - v_{x}(x); 
v_{y}(x + r) - v_{y}(x);  
v_{z}(x + r) - v_{z}(x); 
{\bf B}_{x}(x + r) - {\bf B}_{x}(x);
{\bf B}_{y}(x + r) - {\bf B}_{y}(x); 
{\bf B}_{z}(x + r) - {\bf B}_{z}(x)$; 
where ``$r$'' is the lag or spatial scale. For a given lag this calculation 
is done for each pixel in the three cardinal directions. A normalized 100 bin 
histogram of these values results in our incremental PDF which is then fit 
with the Tsallis function (see Figure \ref{fig:3dhistsall} for an example PDF).

The Tsallis fit parameters are in many ways similar to statistical moments. 
Moments, more specifically the third and fourth order moments, have been 
used to describe the density distributions and have shown sensitivities to 
simulation compressibility (Kowal, Lazarian, \& Beresnyak 2007; Burkhart et al. 2009). The 
first and second order moments simply correspond to the mean and variance of a 
distribution. Skewness, or third order moment, describes the asymmetry of a 
distribution about its mode. Skewness can have positive or negative 
values corresponding to right and left shifts of a distribution respectively. The 
fourth order moment, kurtosis, is a measure of a distribution's peaked or flatness
compared to a Gaussian distribution. Like skewness, kurtosis can have positive or 
negative values corresponding to increased sharpness or flatness. 
In regards to the Tsallis fitting parameters, the $w$
parameter is similar to the second order moment variance while $q$ is
closely analogous to fourth order moment kurtosis. Unlike higher 
order moments, however, the Tsallis fitting parameters are dependent least-squares 
fit coefficients and are more sensitive to subtle changes in the PDF.  

\section{MHD SIMULATIONS}
\label{numericalsetup}
We generate a database of 14 three dimensional numerical simulations (512$^3$
resolution) of isothermal compressible MHD turbulence by using the
Cho \& Lazarian (2002) code and varying the input  values for the sonic
and Alfv\'enic Mach number (See Table \ref{tab:table1}). The sonic Mach number
is defined as ${\cal M}_s \equiv \langle |{\bf v}|/C_s \rangle$, where
$|{\bf v}|$ is the local velocity vector magnitude and $C_s$ is the sound speed. Averaging
is done over the entire simulation. Similarly, the Alfv\'enic Mach number is
${\cal M}_A\equiv \langle |{\bf   v}|/v_A \rangle$, where  $v_A = |{\bf
  B}|/\sqrt{\rho}$ is the Alfv\'enic velocity, $|{\bf B}|$ is the local magnetic 
field vector magnitude, and $\rho$ is density.  Below, we briefly outline the major 
points of the numerical setup (for more details see Cho \& Lazarian 2002). 

The code is  a second-order-accurate hybrid essentially 
non-oscillatory (ENO) scheme which solves
the ideal MHD equations in a periodic box:

\newpage

\begin{eqnarray}
 \frac{\partial \rho}{\partial t} + \nabla \cdot (\rho {\bf v}) = 0, \\
 \frac{\partial \rho {\bf v}}{\partial t} + \nabla \cdot \left[ \rho {\bf v} {\bf v} + \left( p + \frac{B^2}{8 \pi} \right) {\bf I} - \frac{1}{4 \pi}{\bf B}{\bf B} \right] = {\bf f},  \\
 \frac{\partial {\bf B}}{\partial t} - \nabla \times ({\bf v} \times{\bf B}) = 0,
\end{eqnarray}
with zero-divergence condition $\nabla \cdot {\bf B} = 0$, 
and an isothermal equation of state $p = C_s^2 \rho$, where 
$p$ is the gas pressure. On the right-hand side, the source term $\bf{f}$ is a
random large-scale driving force\footnote{${\bf f}= \rho d{\bf v}/dt$}.   Boundary
conditions are periodic. We drive turbulence solenoidally in Fourier space at
wave scale k equal to about 2.5 (2.5 times smaller than L, the size
of the box). This defines the injection scale in our models 
and the driving is done in Fourier space
to minimize the influence of the driving force on the generation of density structures.
The initial density and velocity fields are set to unity.
We do not set the viscosity and diffusion explicitly in our models. 
The scale at which dissipation starts to act is defined by
the numerical diffusivity of the scheme. The ENO-type schemes
are considered to be relatively low diffusion (see Liu
\& Osher 1998; Levy, Puppo, \& Russo 1999). The numerical diffusion depends
not only on the adopted numerical scheme but also on the
smoothness of the solution, so it changes locally in the system.
In addition, it is also a time-varying quantity. All these problems
make its estimation very difficult and incomparable between
different applications. However, the dissipation scales can be estimated
 approximately from the velocity spectra. In the case of
our models we estimate the dissipation scale at $k_{\nu}=30$ pixels.

\begin{deluxetable}{lllllc}
\tablewidth{0pt}
\tablecaption{Simulation Parameters}
\tablehead{
  \colhead{Model} &
  \colhead{B$_{ext}$} &
  \colhead{$\mathcal{M}_{s}$} &
  \colhead{$\mathcal{M}_{A}$} &
  \colhead{Description}
}
\startdata
1  & 0.1 &10.0 & 2.0 & Supersonic \& super-Alfv\'{e}nic\\
2 & 1.0 &10.0 & 0.7 & Supersonic \& sub-Alfv\'{e}nic\\
3  & 0.1 & 7.0 & 2.0 & Supersonic \& super-Alfv\'{e}nic\\
4  & 1.0 & 7.0 & 0.7 & Supersonic \& sub-Alfv\'{e}nic\\
5 & 0.1 & 6.0 & 2.0 & Supersonic \& super-Alfv\'{e}nic\\
6  & 1.0 & 6.0 & 0.7 & Supersonic \& sub-Alfv\'{e}nic\\
7 & 0.1 & 4.0 & 2.0 & Supersonic \& super-Alfv\'{e}nic\\
8 & 1.0 & 4.0 & 0.7 & Supersonic \& sub-Alfv\'{e}nic\\
9 & 0.1 & 3.0 & 2.0 & Supersonic \& super-Alfv\'{e}nic\\
10& 1.0 & 3.0 & 0.7 & Supersonic \& sub-Alfv\'{e}nic\\
11& 0.1 & 0.7 & 2.0 & Subsonic \& super-Alfv\'{e}nic\\
12& 1.0 & 0.7 & 0.7 & Subsonic \& sub-Alfv\'{e}nic\\
13& 0.1 & 0.1 & 2.0 & Subsonic \& super-Alfv\'{e}nic\\
14& 1.0 & 0.1 & 0.7 & Subsonic \& sub-Alfv\'{e}nic\\
\enddata
 \label{tab:table1} 
\end{deluxetable}

As density fluctuations are generated by the interaction of MHD waves,
the time $t$ is in units of the large eddy turnover time 
($\sim L/\delta V$) and the length in units of $L$, the energy injection scale. 
The magnetic field consists of the uniform background
field and a  fluctuating field: ${\bf B}= {\bf B}_\mathrm{ext} + {\bf
  b}$. Initially ${\bf   b}=0$. We divided our models into two groups
corresponding to  sub-Alfv\'enic ($B_\mathrm{ext}=1.0$) and 
super-Alfv\'enic ($B_\mathrm{ext}=0.1$) turbulence.  For each group we
compute several models with different values of  gas pressure (see Table
\ref{tab:table1}).

\begin{figure*}[bt]
  \begin{center}
      \includegraphics[keepaspectratio=true,scale=.70]{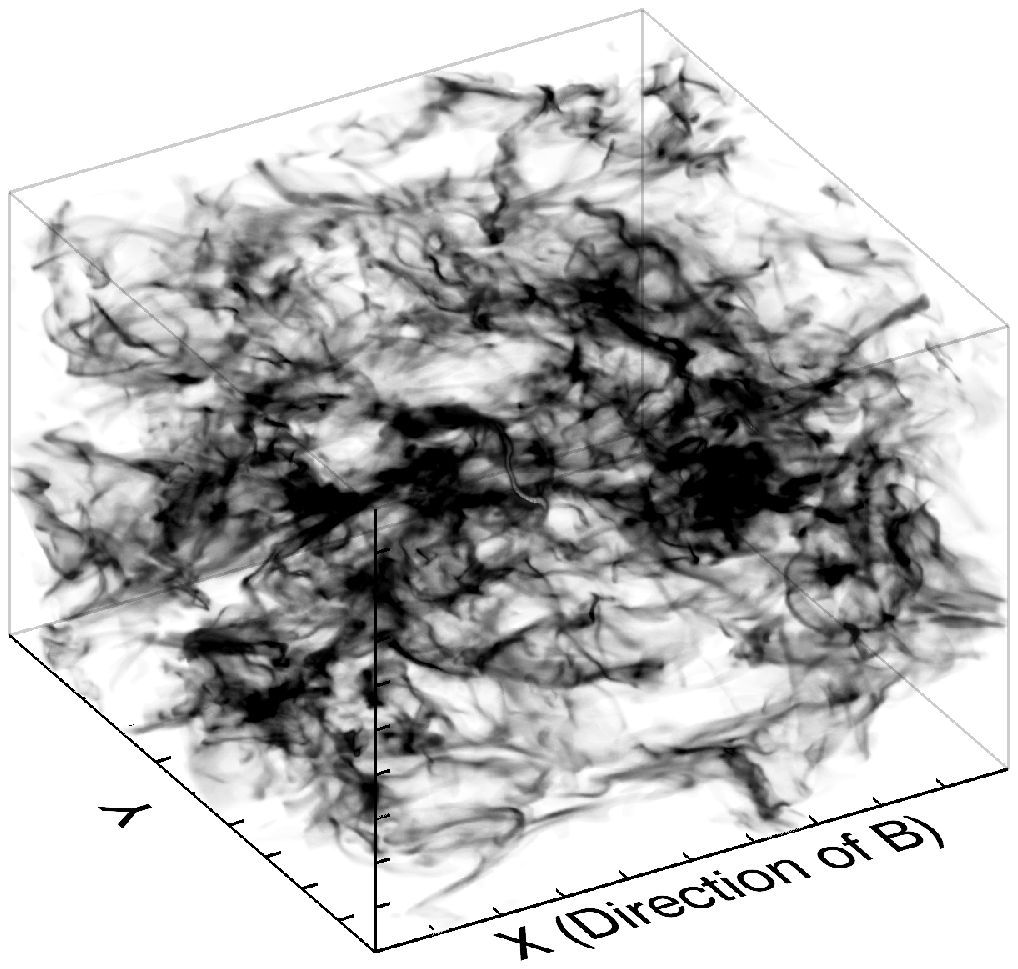}\\
      \includegraphics[keepaspectratio=true,scale=0.35]{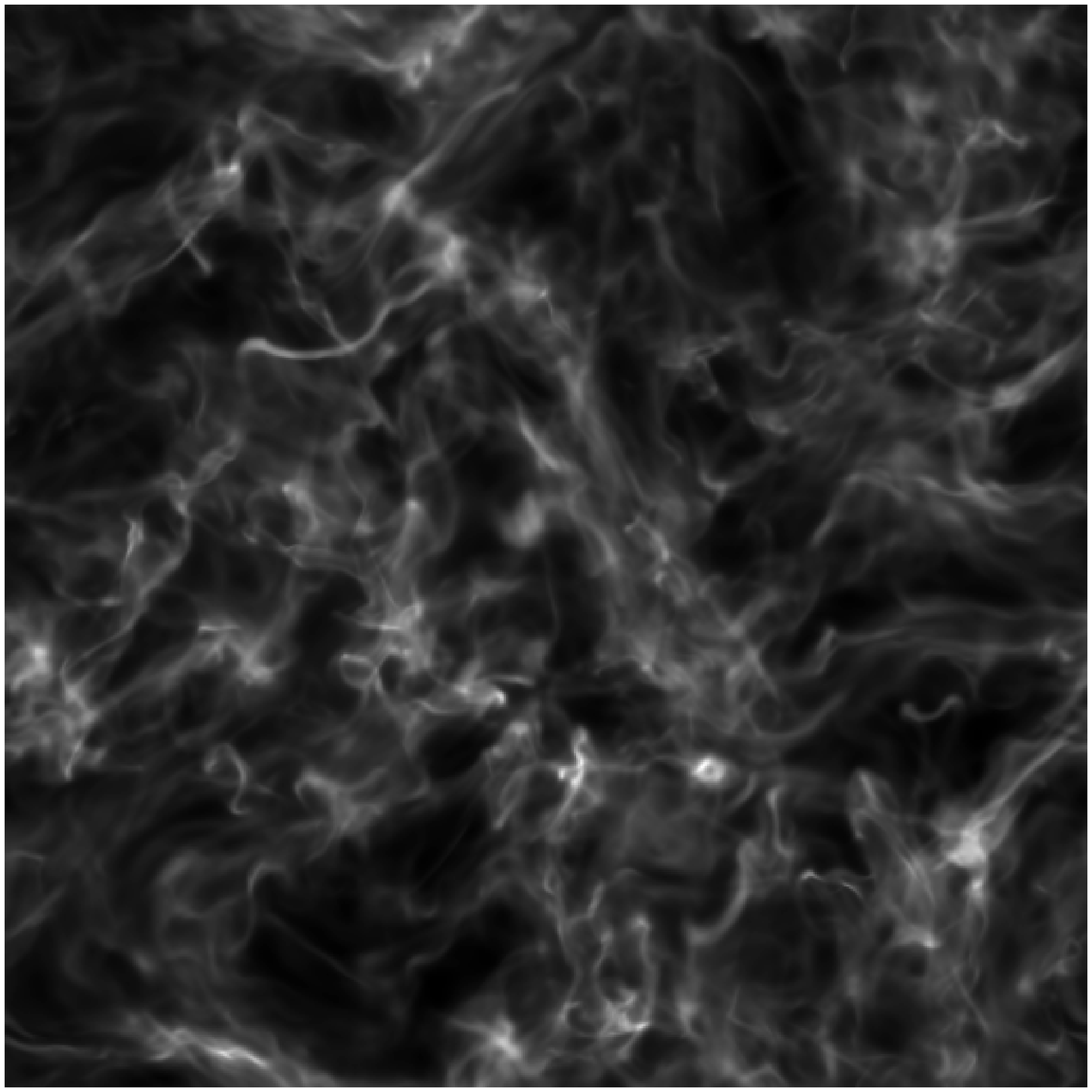}
      \includegraphics[keepaspectratio=true,scale=0.35]{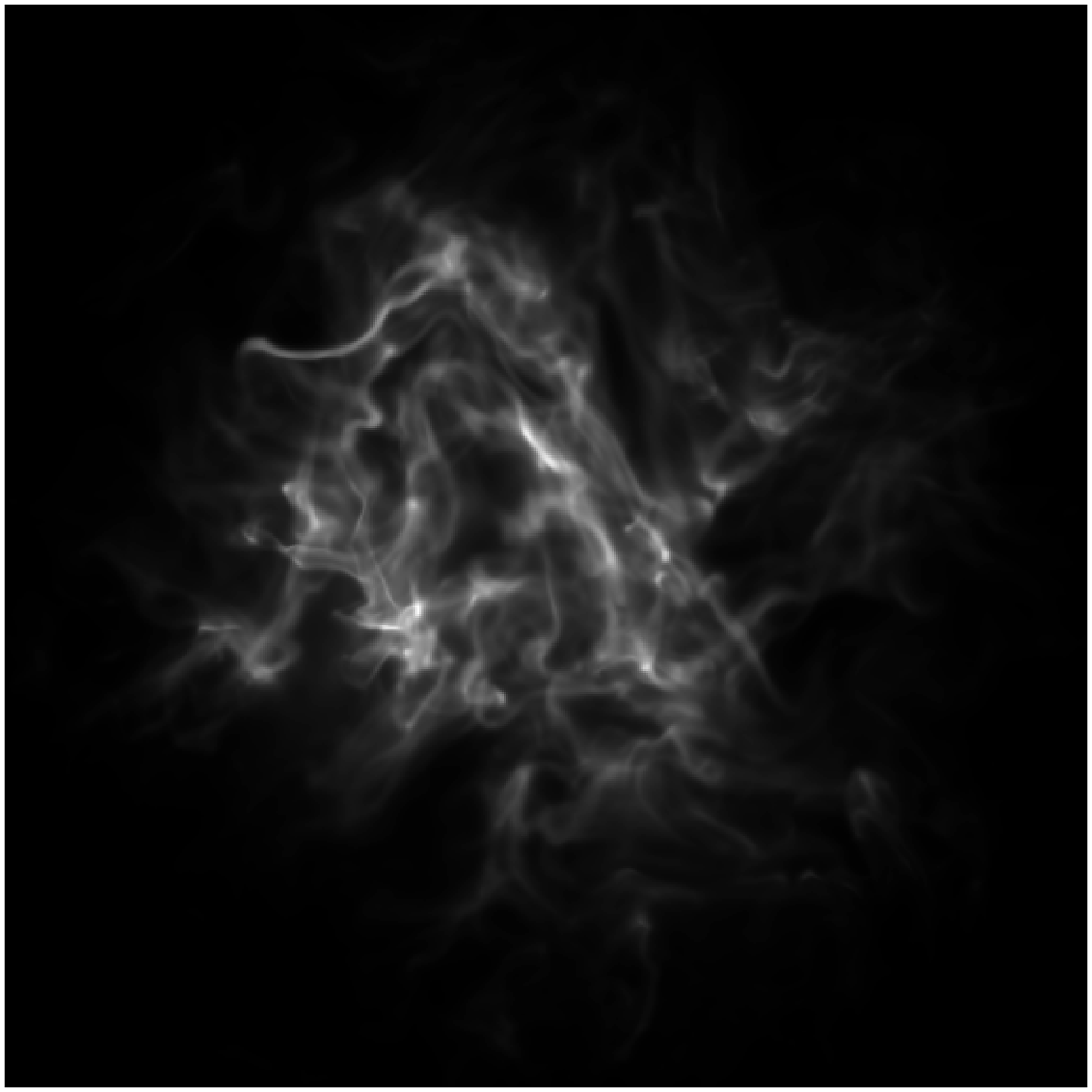}
  \end{center}
  \caption{Top: 3D density rendering of a $512^3$ MHD simulation where
        ${\cal M}_s$=10 (highly turbulent) and ${\cal M}_A$=0.7 (high
        magnetization), Model \#2 from Table \ref{tab:table1}. The mean 
	magnetic field is applied 
	along the x-direction for each simulation. Bottom: Visualizations of 
	compressed 2D column density (left) and cloud bounded column density (right).
	Each is created along the line of sight parallel to the mean magnetic field.} 
  \label{fig:3dpic}
\end{figure*}

The top of Figure \ref{fig:3dpic} presents a 3D density rendering of simulation \#2 (Table
\ref{tab:table1}) where ${\cal M}_s$=10 and ${\cal M}_A$=0.7. The x and y axes
are labeled on the figure with the mean magnetic field ($<${\bf B}$>$)
parallel to the x direction. The bottom displays, for the same simulation, a column 
density (left) and the same column density convolved with a radially decreasing Gaussian function
in order to create the effect of  cloud-like boundaries.  See section 
\ref{observ} for descriptions of column density construction.

\section{Tsallis Fit of Density, Velocity, and Magnetic Field}
\label{3d}
\begin{figure*}[tbh]
  \centering
    \includegraphics[angle=90,keepaspectratio=true,scale=0.6]
    {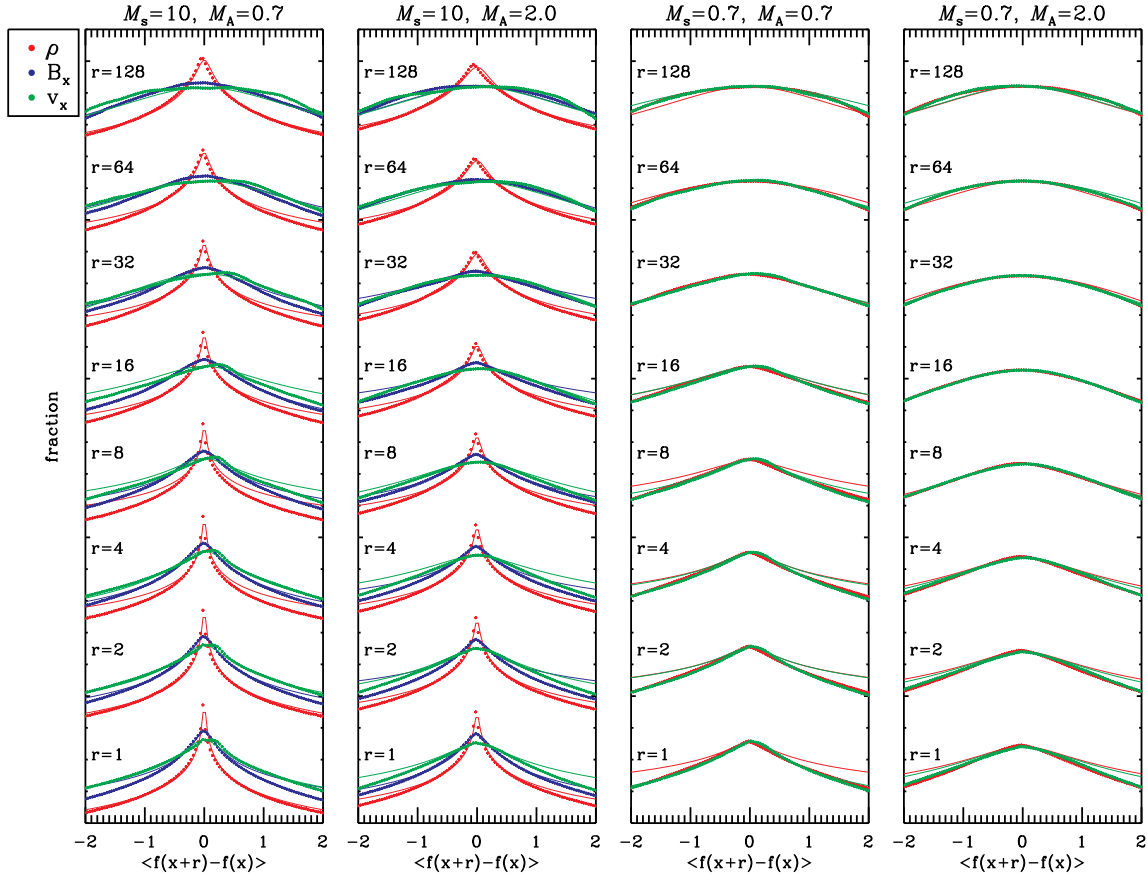}
    \caption{PDFs of 3D density and components of velocity and magnetic
      field parallel to the mean magnetic field 
      (red, green, and blue line respectively). Panels show four different 
      simulations ${\cal M}_s$=10 ${\cal M}_A$=0.7,  ${\cal M}_s$=10 ${\cal M}_A$=2.0, 
      ${\cal M}_s$=0.7 ${\cal M}_A$=0.7, and  ${\cal M}_s$=0.7 ${\cal M}_A$=2.0
      (left to right). PDFs of larger lags are displayed vertically on the 
      same scale for each simulation. Y-axis is logarithmic. Velocity and 
      magnetic field produced similar PDF perpendicular to the mean magnetic 
      field.}
    \label{fig:3dhistsall}
\end{figure*}

For the first portion of our analysis we investigate Tsallis fits of PDFs 
of 3D density and the three directional components of magnetic field and
velocity. We fit the Tsallis distribution to incremental PDFs (see section 
\ref{tsallis}) using the Levenberg-Marquardt algorithm (Levenberg 1944; 
Marquardt 1963), for spatial separations (lag) 1, 2, 4,
8, 16, 32, 64, and 128 pixels (up to 1.5 times smaller then the injection scale). 
Fits and PDFs are shown in Figure \ref{fig:3dhistsall}
(symbols are the data from the simulations, lines represent the Tsallis fit).
Increasing lags are displayed vertically on the same logarithmic 
vertical scale. We only present magnetic and velocity fields directed along the mean
magnetic field. EL10 saw no strong variation with LOS orientation
for magnetic and velocity fields, which we confirm. 
PDFs of perpendicular components are therefore omitted from the figure.  

\begin{figure*}[tb]
  \begin{center}
    \includegraphics[angle=90,keepaspectratio=true,scale=0.5]
      {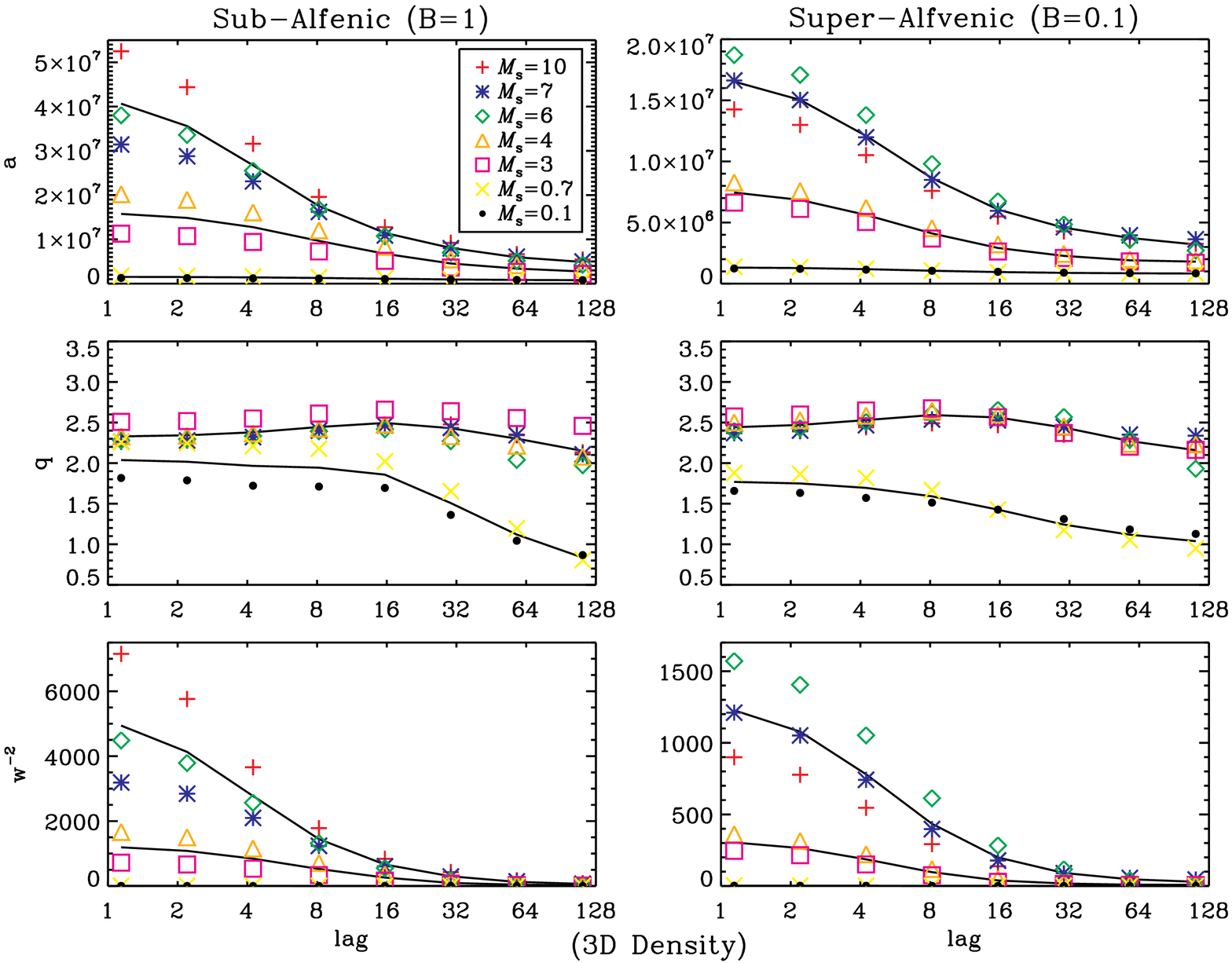}
    \caption{From top to bottom, fit parameters $a$ (amplitude of fit), 
      $q$ (related to fitting tails of PDF), and $w$ (PDF width plotted 
      as $w^{-2}$) are displayed vs spatial lag for all 14  simulations showing 3D 
      density. The left and right columns correspond to sub- 
      and super-Alfv\'enic simulations respectively. Three solid lines are over-plotted 
      for $a$ and $w^{-2}$ averaging the highly supersonic (${\cal M}_s$=10, 7, 6), 
      mid supersonic (${\cal M}_s$=4, 3), and subsonic (${\cal M}_s$=0.7, 0.1) 
      simulations. Two solid lines are over-plotted for the $q$ parameter
      averaging the supersonic (${\cal M}_s$=10, 7, 6, 4, 3) and subsonic
      (${\cal M}_s$=0.7, 0.1) simulations. Parameters $a$ and $w^{-2}$ show 
      significant sensitivities to ${\cal M}_s$ and ${\cal M}_A$ (note 
      scale between left and right panels). Parameter $q$ 
      is only slightly sensitive to the simulation's compressibility. Errors are 
      generally less than 25$\%$, 20$\%$, and 40$\%$ for $a$, $q$, and $w^{-2}$ 
      respectively. Error bars are omitted for clarity.}
    \label{fig:3dfinal}
  \end{center}
\end{figure*}

Figure \ref{fig:3dhistsall} presents PDFs (and their corresponding Tsallis fits)
of 3D density and components of velocity and magnetic field parallel to the mean 
magnetic field (red, green and blue lines respectively). Panels show four different 
simulations ${\cal M}_s$=10 ${\cal M}_A$=0.7,  ${\cal M}_s$=10 ${\cal M}_A$=2.0, 
${\cal M}_s$=0.7 ${\cal M}_A$=0.7, and  ${\cal M}_s$=0.7 ${\cal M}_A$=2.0 (left to 
right). Visual analysis of these figures shows tightly correlated fits
with only slight deviation near the tails of the PDFs (y-axis is logarithmic). 
Three outstanding trends can be
seen across the simulations. First, is the increase in
Gaussianity as ${\cal M}_s$ decreases (turbulence becomes subsonic). 
For supersonic turbulence, density PDFs are highly kurtotic.
This corresponds to a higher probability that 
$\rho(x + r)-\rho(x)=0$ due to shock filaments causing central spikes of in PDFs increasing their
kurtosis (agrees with trends seen in EL10 and Burlaga \& Vi\~{n}as 2004b). 
The second trend is the increase in Gaussianity as the lag
 increases. At low lags density, magnetic field, and velocity have
higher probabilities of being near the mean value (here normalized to zero).
 This type of behavior is congruent with the results of
Falgarone et al.(1994) which analyzed the skewness and kurtosis of PDFs of
varying scales. In the case of subsonic turbulence, the PDFs of density look
very similar to PDFs of velocity and magnetic field for high lags. 
Third, there is an increase in
PDF kurtosis for density, velocity, and magnetic field for simulations of a high magnetic 
field (see the first and second panel of Figure \ref{fig:3dhistsall} for small
lags). Magnetization plays an intimate role in the development of turbulence  
and density enhancements and this affect can be attributed to field freeze-in.
In the following subsections
we will further describe the 3D quantities and their fits individually. 

\subsection{3D Density}
\label{3dd}
From top to bottom, Figure \ref{fig:3dfinal} displays fit parameters $a$, $q$, and $w$ 
(plotted as $w^{-2}$) versus the spatial lag for 3D density for all 14 simulations. 
The figure is separated on the left and right corresponding to sub- and super-Alfv\'enic 
Mach number simulations respectively. The top panels display parameter $a$ 
(corresponding the amplitude) and shows a strong sensitivity to the degree of 
sonic number. To emphasize this fact we break our simulations up into three 
categories; highly supersonic (${\cal M}_s$=10, 7, 6), mildly supersonic (${\cal M}_s$=4, 3), 
and subsonic (${\cal M}_s$=0.7, 0.1). A solid line is plotted 
through each subgroup's average value at each lag. Attention to the difference 
in scale between the right and left panel shows a sensitivity
to ${\cal M}_A$ as well. 

\begin{figure*}[tbh]
  \begin{center}
    \includegraphics[angle=90,keepaspectratio=true,scale=0.5]
    {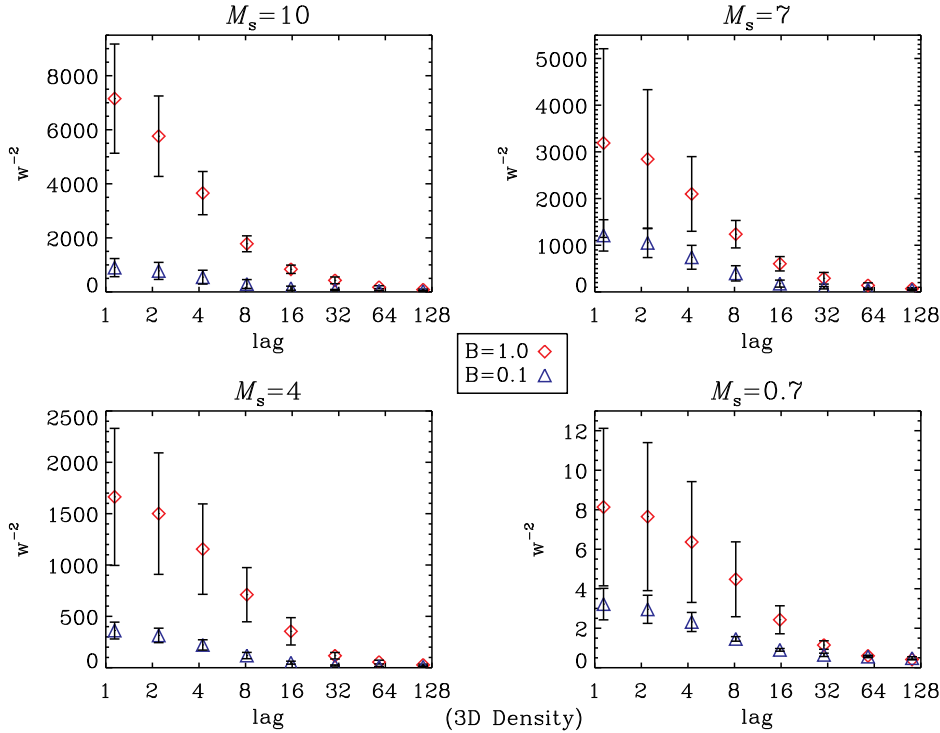}
    \caption{An enlarged version of Figure \ref{fig:3dfinal}'s bottom panel with ${\cal
        M}_s$= 10, 7, 4, 0.7 (top left to bottom right respectively). 
      Sub-Alfv\'enic simulations are denoted with red diamonds while super-Alfv\'enic 
      simulations are denoted with blue triangles. Both super- and sub-sonic simulations show
      larger $w^{-2}$  for high magnetization. Error bars are calculated from the standard 
      deviation of the mean of each ${\cal  M}_s$ group (highly supersonic, mildly supersonic,
      subsonic) at each lag.}  
    \label{fig:3d4}
  \end{center}
\end{figure*}

Errors in the fit parameters are calculated from the standard deviation about each 
subgroup's mean (i.e. highly supersonic, mildly supersonic, subsonic) for each lag. \
Error bars are omitted from 
Figure \ref{fig:3dfinal} for clarity but are displayed in Figure \ref{fig:3d4} for $w^{-2}$. 

The $a$ parameter has a percent error generally $<$ 25$\%$ but reaches a maximum of 56$\%$ for 
simulation ${\cal M}_s$=3.0 ${\cal M}_A$=2.0 (square symbol) at a lag of 1 pixel. The deviation 
from the mean mildly 
supersonic value is not large compared to the supersonic group but the lower numerical 
values produce a higher percent error. This error drops off significantly with increasing lag. 
This is to be expected for simulations since at low lags we are in the range of numerical
dissipation.  Super-Alfv\'enic errors all fall below 20$\%$.

In the middle panels, $q$ (related to the PDF's
tails) shows a slight sensitivity to the simulations compressibility and no
magnetic sensitivity. For $q$ we break the simulations up into only
supersonic (${\cal M}_s$=10, 7, 6, 4, 3) and subsonic (${\cal M}_s$=0.7, 0.1)
and plot a solid line through their average values. The minimal variation in $q$ 
between simulations results in low errors that are below a 20$\%$ for each lag.

Parameter $w$, or width, displayed in the bottom panels of Figure \ref{fig:3dfinal} 
is presented as the fit value $w^{-2}$ as it appears in Equation (1) 
for convenience and clarity of representation.  As with $a$, we plot solid lines 
of the average of highly super-, mildly super-, and subsonic simulations showing the same sensitivity 
to ${\cal M}_s$ although to a lesser degree. Both subsonic simulations lie along the lag (x)
axis. The most outstanding trend seen in $w$ is its sensitivity to ${\cal M}_A$. 
Taking note of the large difference in scales between sub- and
super-Alfv\'enic panels shows elevated values for simulations with high magnetic fields. Figure
\ref{fig:3d4} summarizes this sensitivity displaying the sub- and super-Alfv\'enic versions of
two highly supersonic (${\cal M}_s$=10, 7), one mildly supersonic 
(${\cal M}_s$=4), and one subsonic (${\cal M}_s$=0.7) simulation. For each ${\cal
  M}_s$, the sub-Alfv\'enic ($B_\mathrm{ext}$=1.0) simulation produces a $w^{-2}$ value
$\geq$ 2 times its super-Alfv\'enic counter part at lag = 1 pixel. 

Errors for $w^{-2}$ are the highest of the three parameters due to it large deviation between 
simulations. Generally the errors are $<$ 40$\%$ but peak at 160$\%$ for the ${\cal M}_s$=3.0 
${\cal M}_A$=2.0 simulation (filled circle) at 1 pixel lag. This is mainly due to its $<$ 1 value. 
Even with these significant errors, the differences between sub- and super-Alf\'enic turbulence
is apparent.

Aside from the values of $a$, $q$, and $w$, the general shape of all three
shows trends toward ${\cal M}_s$. Both $a$ and $w^{-2}$ show relatively
consistent values over lag for subsonic simulations while supersonic
simulations show steep decreasing slopes.

\subsection{3D Velocity and Magnetic Field}
\label{3dVB}
The analysis of velocity is more directly related to investigations 
of turbulence than density and is particularly important in regards to 
Solar Wind measurements.  Tsallis provides excellent fits to our velocity PDFs and 
also shows dependencies on Mach numbers, although the Tsallis fit parameters show decreased 
sensitivities to ${\cal  M}_s$ and ${\cal M}_A$ compared to the density analysis.

\begin{figure*}[bh]
  \begin{center}
    \includegraphics[angle=90,keepaspectratio=true,scale=0.45]
    {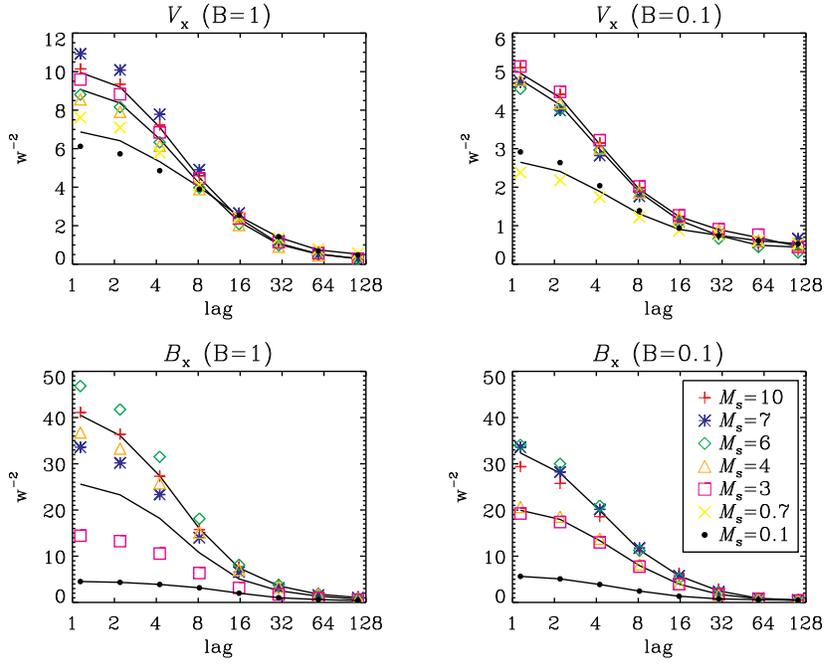}
    \caption{Top: Fit parameter $w^{-2}$ for the velocity component parallel
      to the mean magnetic field for 14 simulations. Sub- and super-Alfv\'enic 
      simulations are presented on the left and right respectively. Three 
      solid lines are over-plotted averaging the super, trans, and subsonic 
      simulations. 
      Bottom: Fit parameter $w^{-2}$ for the magnetic field strength parallel
      to the mean magnetic field for 12  simulations. Three 
      solid lines are over-plotted averaging the super, trans, and subsonic 
      simulations. Errors for $\textbf{V}_x$ are all $<$ 18$\%$. Errors for 
      $\textbf{B}_x$ are all $<$ 20$\%$, excluding mildly subsonic 
      (${\cal  M}_s$=4 \& 3) sub-Alfv\'enic ($B_\mathrm{ext}$=1) simulations.}
    \label{fig:bvx}
  \end{center}
\end{figure*}

The top panels of Figure \ref{fig:bvx} displays 
$w^{-2}$ for sub- and super-Alfv\'enic simulations on the left and right respectively
for the component of velocity along the 
mean magnetic field (x-direction). Solid lines represent the average values of highly super-, 
mildly super-, and sub sonic simulations at each lag (for top and bottom panels).
A small sensitivity to ${\cal M}_s$ can be seen with increased 
values for more turbulent simulations. The sensitivity to ${\cal M}_A$ is most 
predominant parallel to $<${\bf B}$>$ (which is the LOS shown in this figure) 
where there is a near factor of 2 
increase in $w^{-2}$. Perpendicular to $<${\bf B}$>$, the increase is $\le50\%$. 
For sub-Alfv\'enic simulations ($B_\mathrm{ext}$=1) a $\sim$20\% increase in $w^{-2}$ is 
seen for velocity along $<${\bf B}$>$ while no preferred direction is seen for 
super-Alfv\'enic simulations.

We do not show figures for  $a$ and $q$ as they displayed less sensitivity 
to Mach numbers then did  $w^{-2}$.  
Generally for velocity, the $a$ parameter  displays a very small sensitivity to 
compressibility and the values for all simulations span a limited range. 
Compared to 3D density, velocity exhibits a two order of magnitude decrease in 
standard deviation for small lags. A slight increase in $a$ can be seen for
velocity of highly magnetized simulations along the mean magnetic field (x
direction) but no similar relationship is seen for velocity perpendicular to
$<${\bf B}$>$.  Fit parameter $q$ is even less descriptive, with no significant
Mach number dependencies. $w^{-2}$ is the most sensitive to ${\cal
 M}_s$ and ${\cal M}_A$ maintaining the same trends seen in density but on
a fraction of the scale. 

Analysis of directional magnetic field strength  is shown in the bottom of
Figure \ref{fig:bvx} for $w^{-2}$. Tsallis fits the PDFs of magnetic field well.
Parameter $w^{-2}$ displays a sensitivity 
to the ${\cal  M}_s$ number of simulations of a given ${\cal M}_A$ as seen in the bottom 
panels of Figure \ref{fig:bvx} for the magnetic field parallel to $<${\bf B}$>$. 
 In the presented figures $w^{-2}$ has a similar value for high and 
low ${\cal M}_A$ but along other lines of sight the relationship between $w^{-2}$ 
values and increased magnetization generally does not hold.  However, this may
be useful for determining mean field direction in the ISM.  
Similar to velocity, we find that $a$ and $q$ are not as useful as $w$ in terms of describing
Mach numbers with component velocity and magnetic field, and hence we omit the Figures.
Parameter $a$ shows a very small sensitivity to compressibility while $q$ shows no significant 
variation for any simulation.  

Errors analysis for the fit parameters of velocity and magnetic field are carried out 
in the same manner as density (calculating the standard deviation of each ${\cal M}_s$
subgroup at each lag). Velocity along the mean magnetic field 
has fit errors consistently $<$ 18$\%$ for all lags and simulations. Magnetic field along 
the same LOS has similarly low errors ($<$ 20$\%$) with the exclusion of the sub-Alfv\'enic
($B_\mathrm{ext}$=1) ${\cal M}_s$=4 \& 3 simulations (triangles and squares respectively) which 
reach 43$\%$ and 109$\%$ error respectively.
One should keep in mind, however, that these measurements are done for one LOS
as there are significant variations in the trends of velocity and magnetic field depending on 
the orientation.

The behavior of Tsallis fits to PDF increments of density, velocity, and magnetic field
 is in agreement with results found in EL10 at higher resolution with a larger parameter range.
 Our analysis of  spatial variations provides insight into the underlying physics of MHD 
turbulence, where as the analysis of temporal variations of velocity and magnetic field 
in solar wind observations well described the multiphase structure 
of this phenomenon (Burlaga \& Vi\~{n}as 2004a, 2004b, 2005a, 2005b, 2006; 
Burlaga, Ness, \& Acu{\~n}as 2006, 2007, 2009; Burlaga, Vi{\~n}as, \& Wang 2007).
In the latter a time scale, 
$\tau_m$, (opposed to spatial scale $r$) is used to describe incremental 
fluctuations by $(B(t+\tau_m)-B(t))$. 

\begin{figure*}[tb]
  \begin{center}
    \includegraphics[angle=90,keepaspectratio=true,scale=0.45]
    {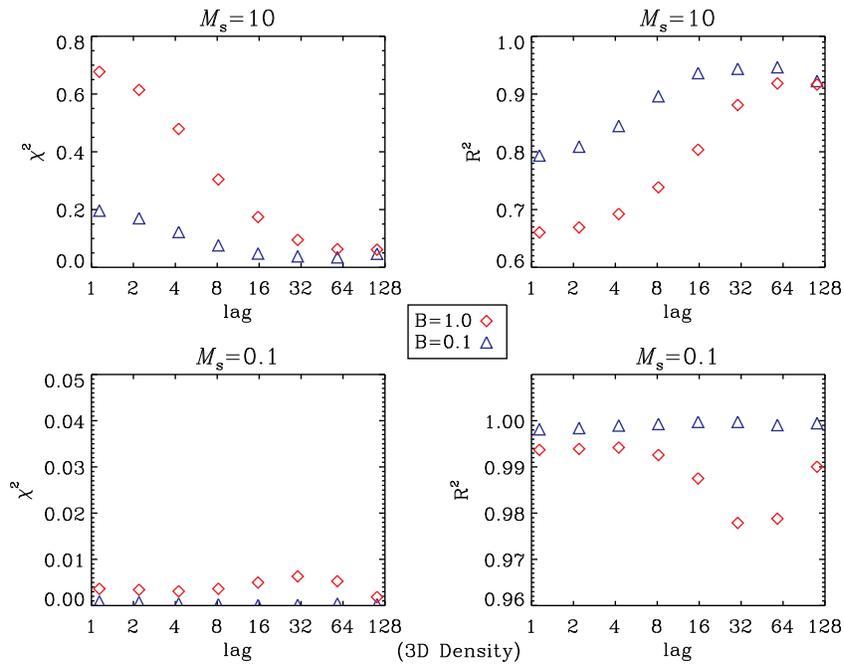}
    \caption{$\chi^2$ and $R^2$ versus lag on the left and right respectively for
      3D density simulations. The sub- and super-Alfv\'enic simulations are over plotted 
      (red diamonds and blue triangles respectively) for
      ${\cal M}_s$=10 (top) and ${\cal M}_A$=0.1 (bottom). Tightest fits are obtained 
      for medium lags for low magnetization and low turbulence. Largest deviations are 
      seen for sub-Alfv\'enic supersonic simulations at small lags (in the dissipation range).}
	\label{fig:chiR}
  \end{center}
\end{figure*}

\subsection{Quality of Fits}
\label{error}
In order to characterize the quality of our fits and reliability of results,
we calculate the Pearson's $\chi^2$ and coefficient of determination, $R^2$,  
for our Tsallis PDF fits presented in Equations (5) and (6) below. 

\begin{eqnarray}
\chi^2 = \sum\limits_{i=1}^n \frac{(y_i - f_i)^2}{f_i}, \\
R^2 = 1-\sum\limits_{i=1}^n \frac{(y_i-f_i)^2}{(y_i-\bar y)^2},
\end{eqnarray}

\begin{figure*}[tbh]
  \begin{center}
    \includegraphics[angle=90,keepaspectratio=true,scale=0.6]
    {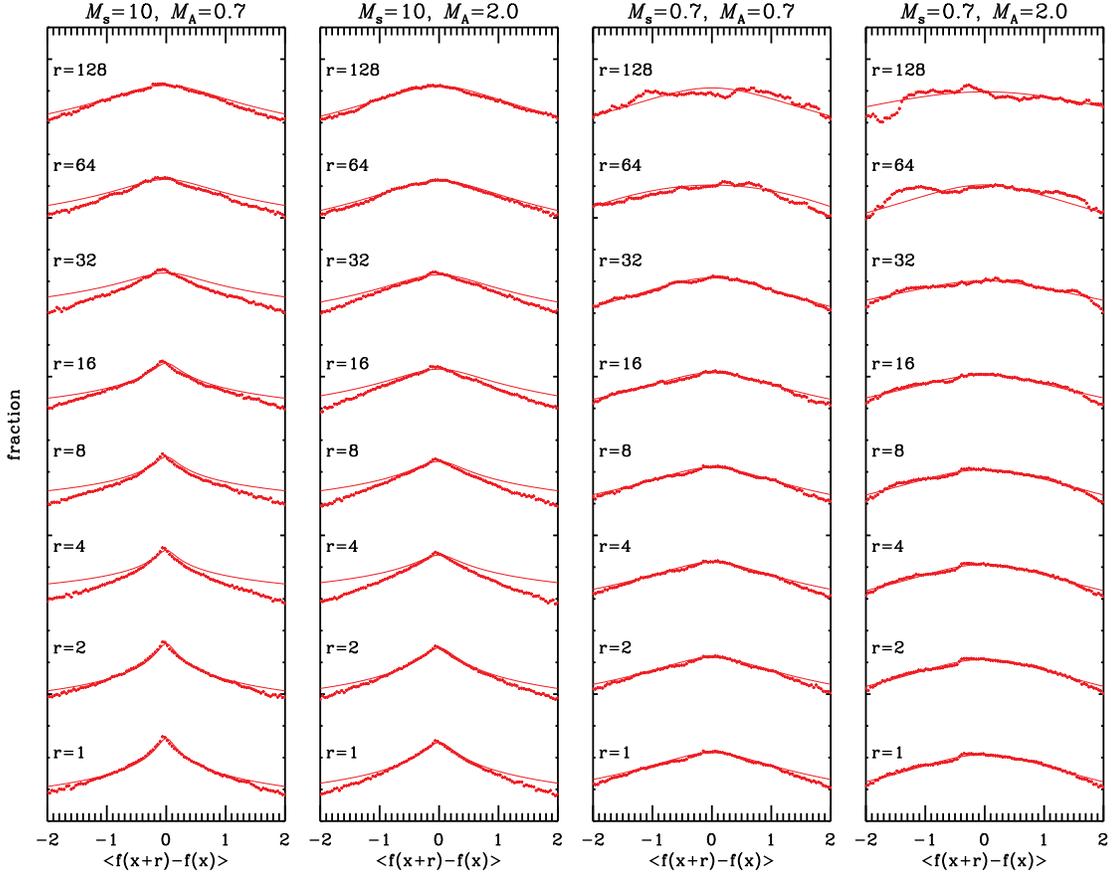}
    \caption{PDFs of column density of four simulations with LOS averaged parallel to the
        mean magnetic field. From left to right, panels
      show ${\cal M}_s$=10 ${\cal M}_A$=0.7,  ${\cal M}_s$=10 ${\cal M}_A$=2.0, 
      ${\cal M}_s$=0.7 ${\cal M}_A$=0.7, and  ${\cal M}_s$=0.7 ${\cal M}_A$=2.0. 
        Column density made
        perpendicular to the mean magnetic field produced similar PDFs. Gaussianity
	of PDFs begins to degrade at high lags for subsonic simulations (top right)}
	\label{fig:cdxhists1}
  \end{center}
\end{figure*}

In these equations, $y$ represents the observed PDF and $f$ is the Tsallis fit.
Each measurement describes the accuracy of the fit. $\chi^2$ approaches 
zero for perfect fits while $R^2$ approaches one. In general, the 
Tsallis function will fit PDFs of density, column density, magnetic field, and 
velocity with a $\chi^2$$\le$0.1 and a $R^2$$\ge$0.85 resulting in tightly 
correlated fits for the 100 degrees of freedom (number of bins in PDF histogram).
Figure \ref{fig:chiR} displays the $\chi^2$ and $R^2$ for the sub- and super-Alfv\'enic 
counterparts  of our most turbulent (top) and least turbulent 3D density 
simulations (bottom). From these plots we find that the best fits are seen for 
super-Alfv\'enic, subsonic simulations for lags greater than 16 pixels. 
Due to the kurtotic nature of high turbulence, low lag PDFs are difficult 
for Tsallis to fit at the central peak (see Figure \ref{fig:3dhistsall}, right 
panel, bottom). In these extreme cases (${\cal M}_s$=10, ${\cal M}_A$=0.7) values 
of $\chi^2$=0.67 and $R^2$=0.61 are obtained. 

The low lag regimes are in the dissipation range of 
the turbulence, and Tsallis seems to have a difficult time fitting turbulence on scales sampling either
the dissipation range or the injection scale. Despite these difficulties at low (or high) lag, Tsallis still can prove itself
useful for characterizing turbulence.  For example,  if one is in the dissipation range of turbulence (i.e. at scales similar to our lags
less then 32 pixels), then these ``goodness of fit tests'' can also be used to 
determine what type of turbulence is present.  Subsonic  and super-Alfv\'enic turbulence both are better fit with the Tsallis distribution at low lags then
supersonic or sub-Alfv\'enic turbulence.  At higher lags the fit quality converges as we enter the inertial range of turbulence.  
These fit tests should not only be preformed to test fit quality but could also be used as an additional test of the Mach number range in a given data set.

\begin{figure*}[tbh]
  \begin{center}
    \includegraphics[angle=90,keepaspectratio=true,scale=0.5]
    {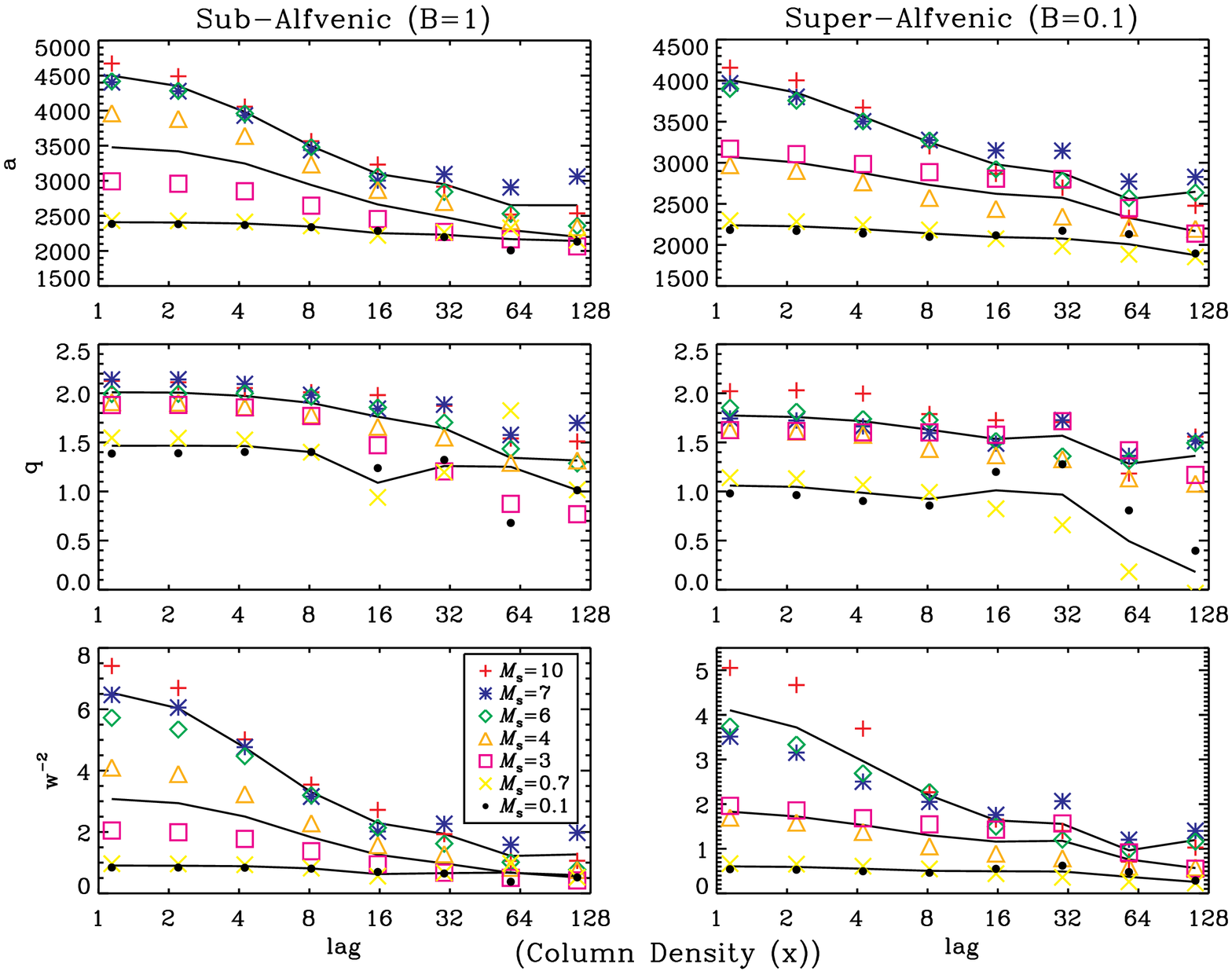}
    \caption{From top to bottom fit parameters $a$ (amplitude of fit), 
      $q$ (related to fitting tails of PDF), and $w$ (PDF width plotted 
      as $w^{-2}$) are displayed vs spatial lag for all 14 simulations. Column 
      density are LOS averaged parallel to the mean magnetic field. 
      The left and right columns correspond to sub- and super-Alfv\'enic simulations 
      respectively. Three solid lines are over-plotted 
      for $a$ and $w^{-2}$ averaging the highly supersonic (${\cal M}_s$=10, 7, 6), 
      transonic (${\cal M}_s$=4, 3), and subsonic (${\cal M}_s$=0.7, 0.1) 
      simulations. Two solid lines are over-plotted for the $q$ parameter
      averaging the supersonic (${\cal M}_s$=10, 7, 6, 4, 3) and subsonic
      (${\cal M}_s$=0.7, 0.1) simulations. Parameters $a$ and $w^{-2}$ show 
      sensitivities to ${\cal M}_s$ and ${\cal M}_A$ (note 
      scale between left and right columns). Parameter $q$ 
      is only slightly sensitive to the simulation's compressibility. Lags from 2 to 
      32 pixels provide the most consistent fit parameter errors where $a$, $q$, 
      and $w^{-2}$ maintain errors $<$ 22$\%$, 35$\%$, and 24$\%$ respectively.} 
    \label{fig:cdxall}
  \end{center}
\end{figure*}

An alternative source of potential error in our analysis is the quality of PDFs 
at large lags. Not only are these lags on the upward scale of the inertial range, 
but also are subjected to degraded resolution.  These resolution effects can be seen particularly in the case
of the column density, which are 2D quantities discussed in the next section. 
The inspection of Figure \ref{fig:cdxall} shows fits becoming less tight at large lags.  This manifests as ``jumps''
in the fits parameters  which can be seen in the subsonic case in Figure \ref{fig:cdx4}. This is 
the most extreme example of this trend but it is seen to some degree for every 
simulation at a large enough lag. In these cases the $\chi^2$ and $R^2$ remain constrained to the
values stated above. We conclude that, for this analysis, our 
results are most stable for lags in the inertial range of turbulence and for lags at least 8 times smaller then the box size to provide
enough sampling.  This 
trend is consistent with the results seen using higher order moments to 
analyze the incremental PDFs. 

\newpage

\section{Tsallis Fit of 2D Column Density and PPV data}
\label{observ}
In section \ref{3d} we confirmed the results of EL10 using higher resolution
simulations and a substantially larger parameter range.  In addition, we
demonstrate the sensitivity the $w$ (width) fit parameter has toward the
Alfv\'enic Mach number. However, observations of the ISM do not provide
direct 3D information of density.  Combining density and velocity along the
line-of-sight (LOS) provides a 3D position-position-velocity (PPV) cube,
however these types of data cubes can almost never be reliably interpreted
as having a one-to-one correspondence with an actual 3D volume density.
Considering the difficulties of obtaining direct 3D ISM information, a
study of  Tsallis statistics on observables such as column density and PPV
data is necessary.

\subsection{Column Density}
\label{coldn}

We create synthetic 2D column density maps perpendicular and parallel to
the mean magnetic field for all 14 of our simulations by averaging the 3D
density cubes along a given line of sight. We assume  the emitting gas is
optically thin and that the emissivity is linearly proportional to density
(such as in the case of HI). An example is presented in Figure
\ref{fig:3dpic} (bottom left). Using the same method described in the section
\ref{tsallis}, PDFs are created using spatial lags (increments) in \emph{two}
directions which are then fit using the Tsallis function. 

Figure
\ref{fig:cdxhists1} displays the distributions and fits
for 4 simulated column densities created parallel to the mean magnetic field.
The 4 simulations we show here are divided into panels:
${\cal M}_s$=10 ${\cal M}_A$=0.7,  ${\cal M}_s$=10 ${\cal M}_A$=2.0, 
${\cal M}_s$=0.7 ${\cal M}_A$=0.7, and  ${\cal M}_s$=0.7 ${\cal M}_A$=2.0
(left to right). This is analogous to the arrangement in Figure 
\ref{fig:3dhistsall}. While there is an 
overall decrease in kurtosis of the column density PDFs compared to 3D density, 
the same three trends are still apparent. PDFs become more Gaussian with 
decreased ${\cal M}_s$, Gaussianity increases with lag for supersonic cases, and 
Gaussianity increases with ${\cal M}_A$. A trend not seen in the 3D case is that 
subsonic PDFs become less smooth at high lags. This is due to the decrease in 
resolution going from 3D to 2D.  EL10 also observed this trend
with their subsonic PDFs.

\begin{figure*}[tbh]
  \begin{center}
    \includegraphics[angle=90,keepaspectratio=true,scale=0.5]
    {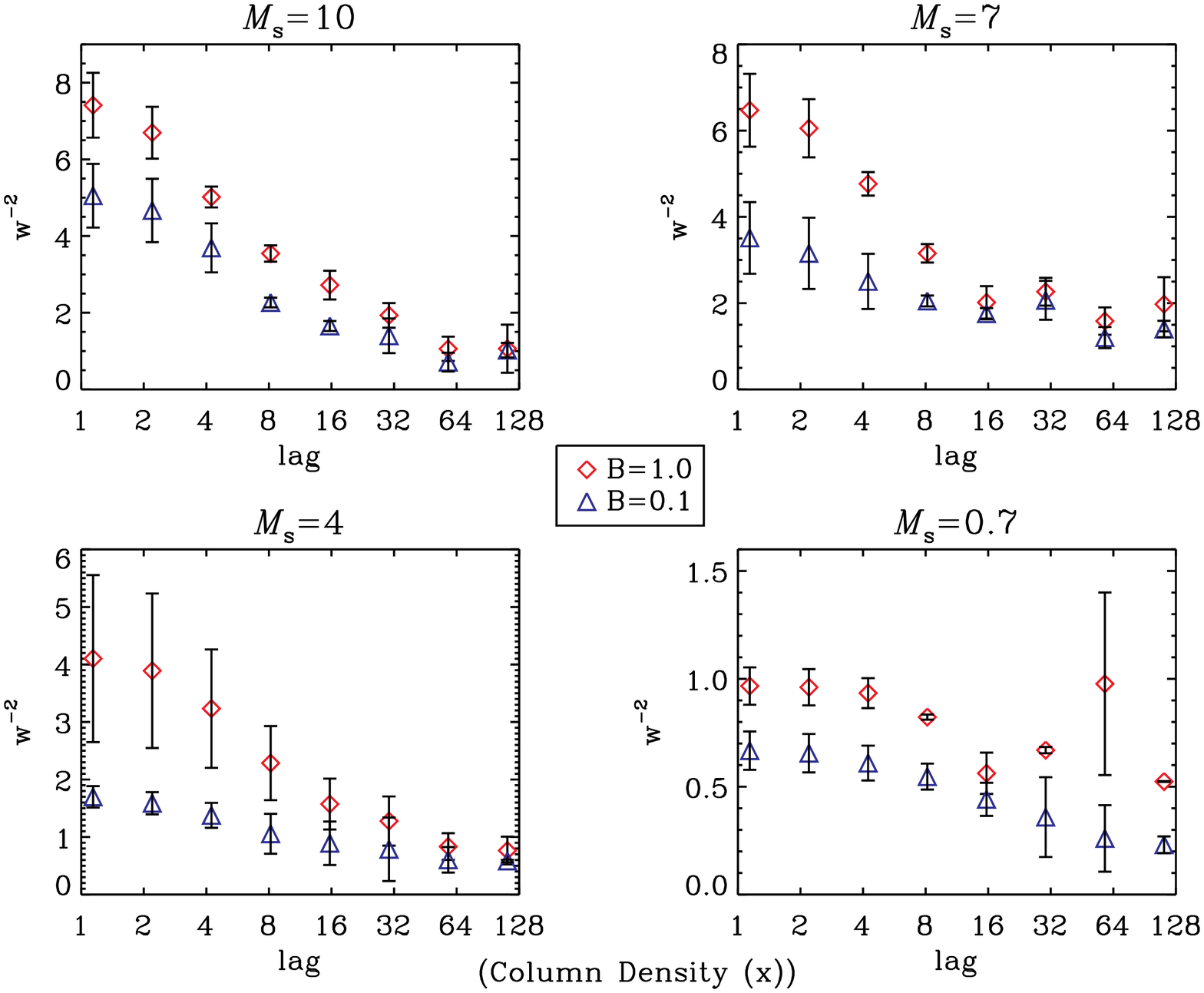}
    \caption{An enlarged version Figure \ref{fig:cdxall}'s bottom panel with ${\cal M}_s$= 10,
      7, 4, 0.7 (top left to bottom right respectively). High magnetization is
      denoted with red diamonds and low magnetization is denoted with  blue
      triangles. Both super- and subsonic simulations show larger $w^{-2}$
      for high magnetization. Error bars are calculated from the standard 
      deviation of the mean of each ${\cal  M}_s$ group (highly supersonic, mildly supersonic,
      subsonic) at each lag.} 
    \label{fig:cdx4}
  \end{center}
\end{figure*}

Figure \ref{fig:cdxall} displays the fit parameters $a$, 
$q$, and $w^{-2}$ from top to bottom for all 14 column density
simulations created along the x direction (parallel to $<${\bf B}$>$). 
Each parameter is divided into two panels with 
sub-Alfv\'enic  and super-Alfv\'enic simulations on the left and right respectively. The
$a$ parameter (top panels) is presented with three over-plotted
lines, averaging the super-, mildly super-, and subsonic
simulations. Again, a strong sensitivity to ${\cal M}_s$ is seen both in the
value and the shape of the fit over spatial scale. The sensitivity to ${\cal
  M}_A$ is less prevalent than in 3D density.

Errors for column density fit parameters are calculated in the same manner as 3D 
density by measuring the standard deviation of each ${\cal M}_s$ subgroup at each
lag. The $a$ parameter achieves $<$ 23$\%$ errors for each lag and simulation 
reaching the lowest errors ($<$ 5$\%$) at lags between 16 and 64 pixels. As in density,
error bars are excluded from Figure \ref{fig:cdxall} for clarity but displayed in 
Figure \ref{fig:cdx4} for $w^{-2}$.

Parameter $q$ (middle panels) 
is presented with averaging super- and subsonic lines showing
a slight sensitivity to ${\cal M}_s$ and no coherent sensitivity to ${\cal
  M}_A$. Errors for $q$ are generally less than 30$\%$ but begin to reach 
percent errors greater than 100$\%$ for subsonic simulations at lags greater 
than 34 pixels due to random variations. Inspection of the right panel of Figure
\ref{fig:cdxhists1} shows that this is consistent with when the PDF start to become 
nonuniform.

The $w$ parameter (bottom panels), here plotted as $w^{-2}$, 
shows both a strong sensitivity to ${\cal M}_s$ and ${\cal M}_A$. Highly super-,
mildly super-, and subsonic averaging lines are over-plotted which emphasize that values of $w^{-2}$ are affected by ${\cal
  M}_s$. Errors for $w^{-2}$ are $<$ 25$\%$ for lags of 2 to 32 pixels. Maximum 
errors reach values of 70$\%$ at a 1 pixel lag and 60$\%$ error at lags greater than 
64 pixels. Interestingly, as in EL10, subsonic column densities shows a more random behavior
at larger lags due to low resolution and low density contrasts.

Figure \ref{fig:cdx4} summarizes $w$'s Alfv\'enic sensitivity by
displaying the sub- and super-Alfv\'enic versions of two highly supersonic (${\cal M}_s$=10,
7), one mildly supersonic (${\cal M}_s$=4), and one subsonic (${\cal M}_s$=0.7)
simulation. In each case, the simulation having the higher magnetic field
results in a higher $w^{-2}$ value. Although the degree to which $w^{-2}$ is
elevated is less than for 3D density, a strong direct correlation remains.
Subsonic trends become less smooth at lag=32 pixels, which is $\approx$ 6.4 times smaller
then the injection scale.  We may conclude that the column density has a stronger
 dependency on lag then in the 3D cases, especially for subsonic turbulence.

We analyze fit parameters from column densities created along the y and z axes
(i.e. perpendicular to the mean magnetic field) and while
there were slight variations in the fits, sensitivities to Alfv\'en and
sonic Mach numbers were still observed. No discernible trend to detect
magnetic LOS orientation was seen. 

\begin{figure*}[tbh]
  \begin{center}
    \includegraphics[angle=90,keepaspectratio=true,scale=0.5]
    {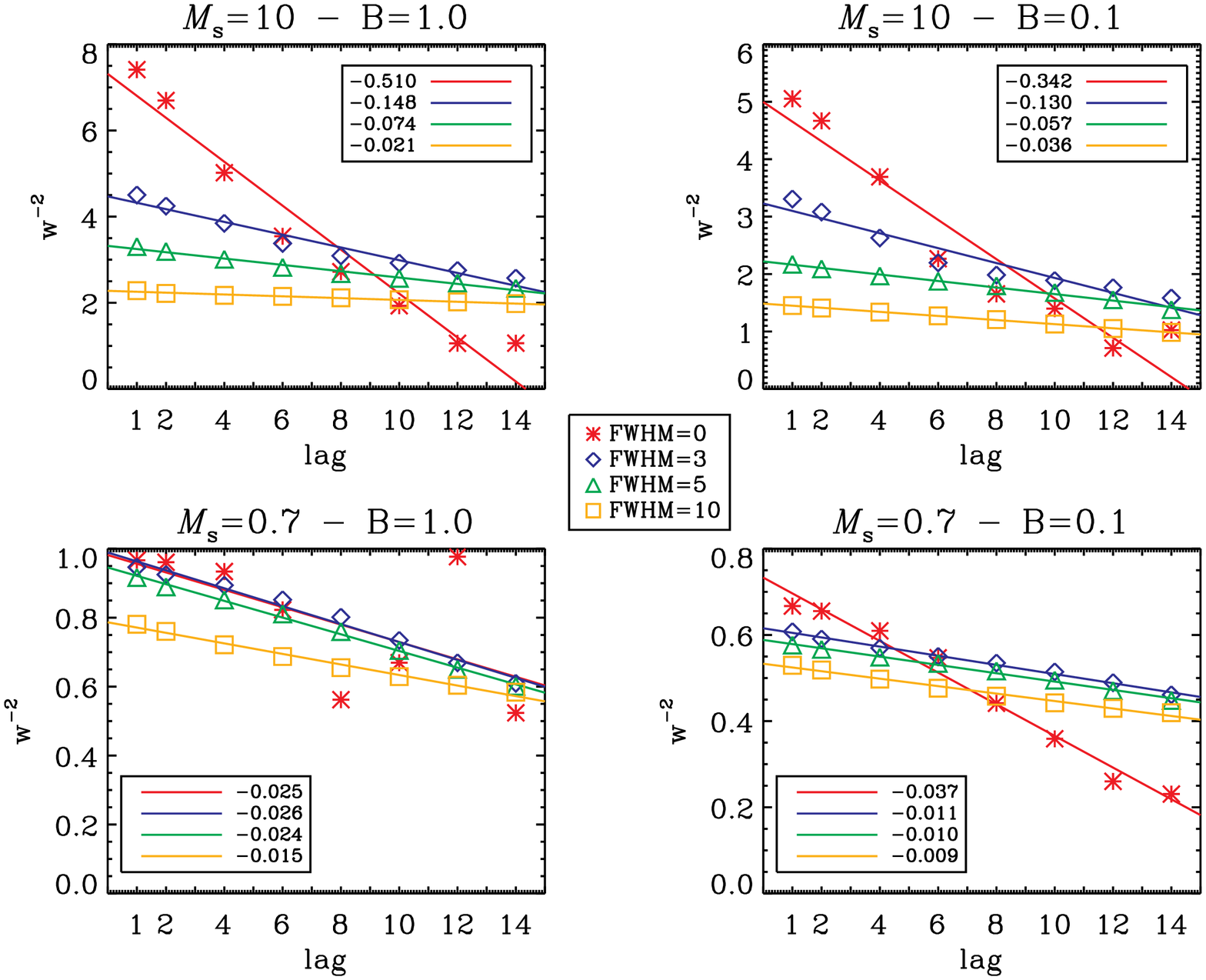}
    \caption{Fit parameter $w^{-2}$ for  ${\cal M}_s$= 10 (top), and 0.7
      (bottom) with sub- and super-Alfv\'enic simulations on the left and right
      respectively. The original and smoothed data are over-plotted. Least squares 
      linear fits are over plotted in the same color.  The enclosed legends displays
      the slope of the respective linear fits. The central legend provide the
      FWHM of smoothing (FWHM=0 is unsmoothed). Increased smoothing results in
      decreased values and shallower slopes.} 
    \label{fig:lag1}
  \end{center}
\end{figure*}

\begin{figure}[tbh]
  \begin{center}
    \includegraphics[angle=90,keepaspectratio=true,scale=0.3]
    {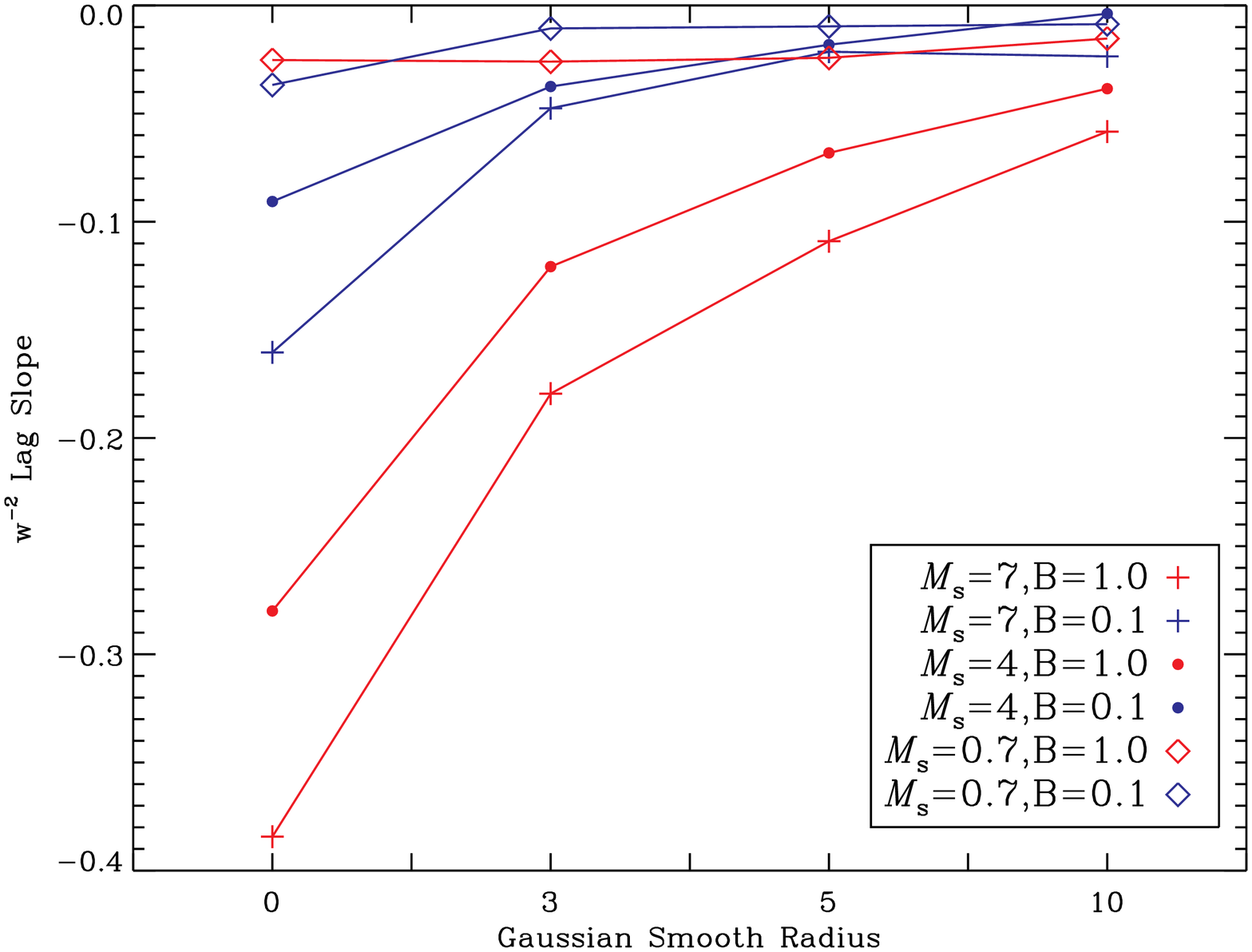}
    \caption{Slope of $w^{-2}$ (width) for the sub- and super-Alfv\'enic versions of the
      ${\cal M}_s$=7, 4, and 0.7 simulations plotted against the smoothing degree in FWHM
      size. High and low magnetization are presented in red and blue
      respectively. Supersonic simulations are more dramatically affected than
      subsonic as turbulent density enhancements are smoothed out.}
    \label{fig:mvsmooth}
  \end{center}
\end{figure}

\subsubsection{Effects of Smoothing}
\label{smooth}
While the analysis simulated column density maps shows the
strength of Tsallis statistics in describing turbulent characteristics, analogous observational 
column density data do not have pencil thin beam resolution. To address this issue, 
we experiment with smoothing our column densities to different degrees to determine 
the role degradation of resolution plays in distributions and fit
sensitivities. Using a Gaussian smoothing kernel, we degrade our column
density simulations with a full-width-half-maximum (FWHM) of 3, 5, and 10
pixels.

In an effort to characterize the effects of smoothing we increase our lag
resolution by creating the same number of distributions over a smaller
range of lags. Figure \ref{fig:lag1} presents the the original and smoothed
fit parameter $w^{-2}$ for sub- and super-Alfv\'enic simulations versions of simulations ${\cal
  M}_s$=10 and 0.7 from top left to bottom right respectively. While we investigated the effects of smoothing
on all three fit parameters, we only present $w^{-2}$ as the effects on $q$ and $a$ are similar.
In order to characterize how the slope of the
Tsallis parameters vs. lag change with increasing smoothing a least squares
linear fit is applied to each scenario with the slopes displayed in the
respective legends. The red stars denote the non-smoothed case for comparison. 
From this figure it is clear that increased smoothing
lowers the values of $w^{-2}$ for each lag and its overall slope. 
These effects are present across all simulations. 
 While the supersonic (top row) simulations show smooth monotonic trends for
both smoothed and pencil beam data, subsonic simulations (bottom row) shows very bumpy 
trends for the case where no smoothing is introduced (red stars and line)
especially as lag increases.  However, as we introduce smoothing, the trends 
become highly monotonic, even for the case of FWHM=3 pixels. Applying Tsallis 
fits to incremental PDFs while varying the level of smoothing will act as an 
additional tool to characterize turbulent parameters through the overall 
change in slope.

Figure \ref{fig:mvsmooth} provides a summary of the effect smoothing has on
the slope of $w^{-2}$ by plotting it against the smoothing degree. 
Simulations of the same sonic number are shown with the same symbol while 
the color corresponds to the previously used color scheme (red being  $B_\mathrm{ext}$=1, 
sub-Alfv\'enic and blue being $B_\mathrm{ext}$=0.1, super-Alfv\'enic).
We see that increased smoothing flattens the $w^{-2}$-lag slope. A 
sensitivity to the ${\cal M}_A$ is still present in both slope and value in
every smoothing cases (note scale between sub- and super-Alfv\'enic simulations in Figure 
\ref{fig:lag1}). High sonic Mach number simulations show slopes that are the most altered 
as smoothing is increased. This is due to shock density enhancements being smoothed out. 
However, the supersonic cases show larger variation with differing ${\cal M}_A$ then 
their subsonic counterparts. This leads us to believe that one can more easily tell the 
magnetization strength of supersonic gas from Tsallis fits. Increasing the smoothing does 
not affect the subsonic cases, as these simulations already have Gaussian distributions and have no peak density enhancements
from shocks to smooth out.

Parameter $a$ was affected by smoothing in the exact same way $w^{-2}$. Values and 
slopes decreased with increased smoothing. Plots are excluded for brevity but both $a$ and
 $w$ could be explored with increasing smoothing to put estimates on turbulence. 
The $q$ parameter saw little deviations due to smoothing and a slightly weakened
${\cal M}_s$ sensitivity.

Fit parameter errors for our smoothing discussing can be considered to decrease with 
smoothing from their unsmoothed values (quoted in the previous section) as differences 
in ${\cal M}_s$ sub groups of simulations become washed out in parameter space.

\subsubsection{Artificial Noise}
\label{noise}
While analysis using Tsallis statistics is very adept at providing turbulence 
parameters when smoothing 
is applied to column density maps, more pressing maybe the issue of noise.  
One may not expect noise to affect results at larger lags however, noise may make it 
challenging to distinguish trends on smaller scales where the difference between 
our simulations is  prominent. It is our intent in this subsection to test the 
effectiveness of this statistical tool with the addition of noise, determine our confidence 
range between trends with noise and with no noise, and explore which 
lags are most affected by the addition of noise. In order to achieve this end 
we add random Gaussian noise to our column density maps.  We do this by setting a 
given average signal-to-noise (SNR) ratio and scaling the power of the noise and 
signal to match this SNR.  We look at column density integrated parallel to the mean 
magnetic field with average SNR of 400, 200, 100, 50, 20 and unity.  

\begin{figure*}[bh]
  \begin{center}
    \includegraphics[angle=90,keepaspectratio=true,scale=0.50]
		    {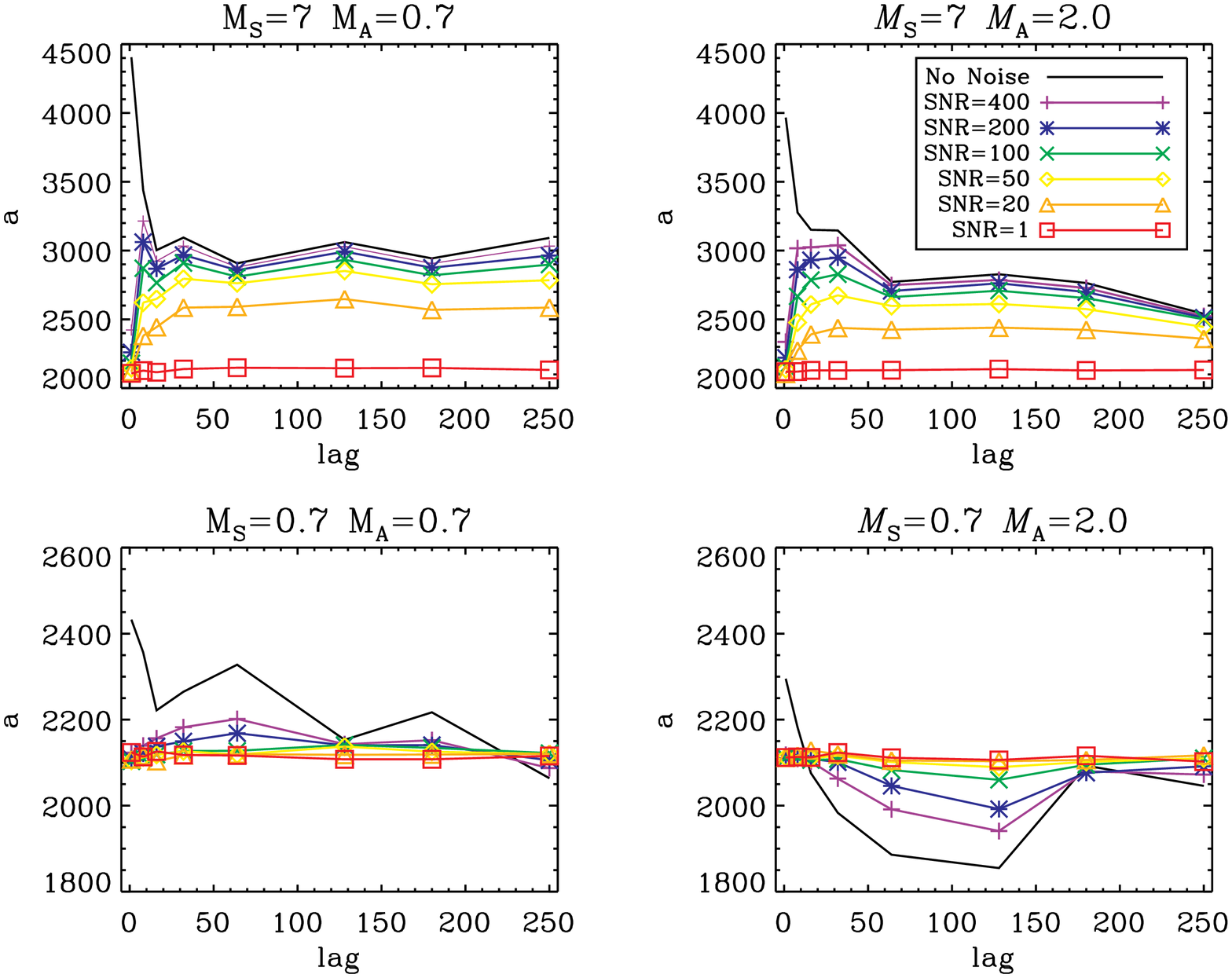}
    \caption{Fit parameter $a$ for 4 simulations (sub and super-Alfv\'enic of simulations 
      ${\cal M}_s$=7, 0.7) with varying levels of noise (see legend). Small amounts of 
      noise drastically 
      lower small lags values and has an overall smoothing affect. Increasing noise lowers $a$'s 
      sensitivity to ${\cal M}_s$ and ${\cal M}_A$.}
    \label{fig:noisea}
  \end{center}
\end{figure*}

\begin{figure*}[tb]
  \begin{center}
    \includegraphics[angle=90,keepaspectratio=true,scale=0.50]
		    {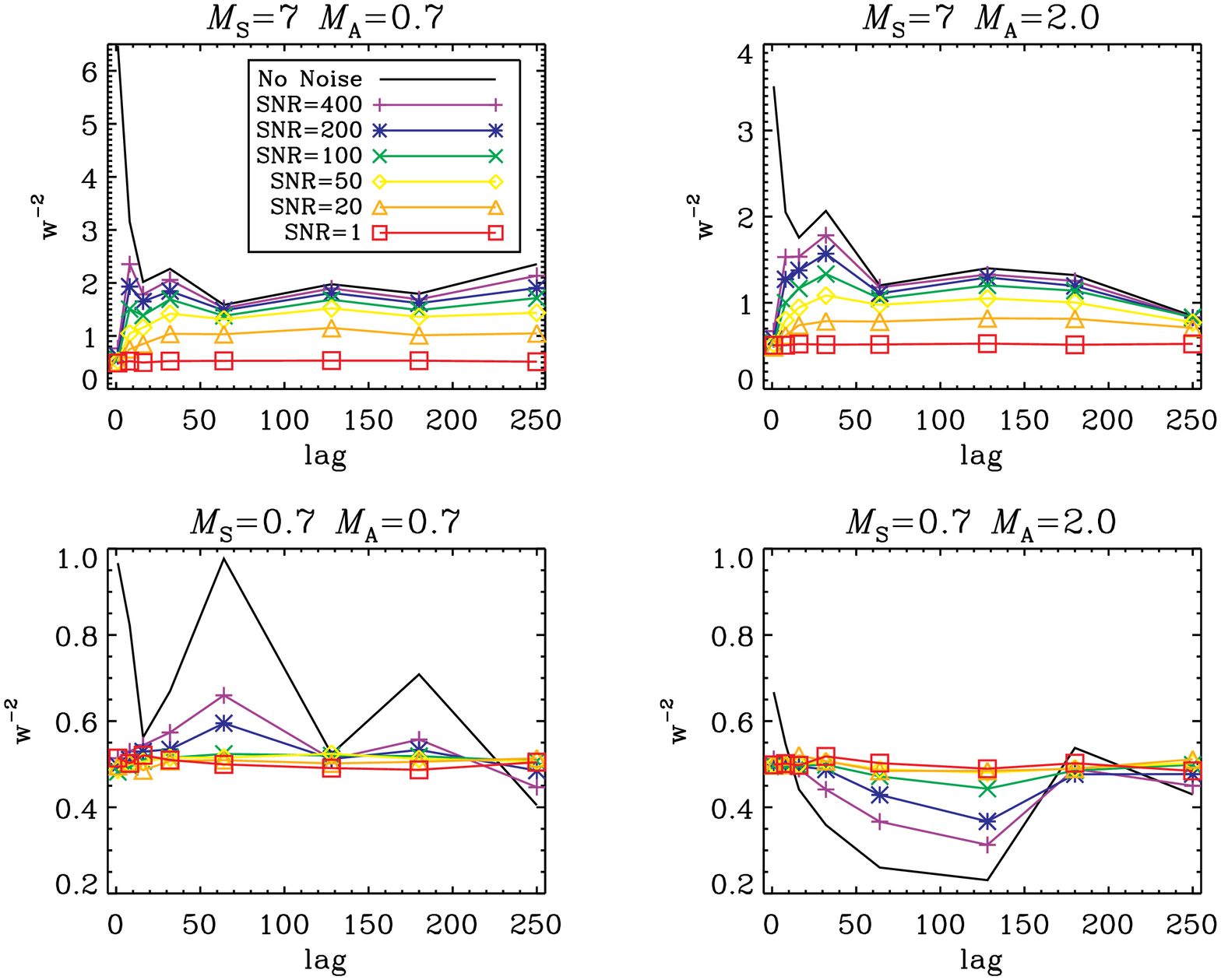}
    \caption{Fit parameter $w^{-2}$ for 4 simulations (sub and super-Alfv\'enic of simulations 
      ${\cal M}_s$=7, 0.7) with varying levels of noise (see legend). Small amounts of noise 
      drastically lower small lags values and has an overall smoothing affect. Increasing 
      noise lowers $w^{-2}$'s sensitivity to ${\cal M}_s$ and ${\cal M}_A$.}
    \label{fig:noisew}
  \end{center}
\end{figure*}

The addition of noise on the column density PDFs has a general smoothing effect 
toward Gaussianity (PDFs not shown). This is to be expected as the noise added
to the column density is Gaussian white noise.  Figures \ref{fig:noisea} and 
\ref{fig:noisew} illustrate the effects of noise for the parameters $a$ and $w^{-2}$ 
respectively for four simulations with given sonic Mach numbers of ${\cal M}_s$= 7.0 
and 0.7 and Alfv\'enic number ${\cal M}_A$= 2.0 and 0.7. From these figures it is 
clear that the smaller spacial lags (lags = 1 - 32 pixels) are highly affected by the 
addition of noise. The noise injected simulations (lines with colors and symbols) do 
not converge to their no-noise counter part (shown as a solid black line) until lag 
$\sim$20 pixels.  At around lag 20, simulations with SNR greater than 50 show similar 
shaped curves to the no-noise case. As the SNR decreases, the values of $w$ and $a$  both 
decrease for the supersonic case.  The subsonic case shows a more pronounced non-monotonic
behavior.  In all cases above SNR of 20, the sub-Alfv\'enic supersonic turbulence shows 
heightened values of $a$ and $w^{-2}$.  The sonic number between these
two simulations can be determined within 1.5 $\sigma$ confidence at SNR 20 and above. 
At SNR of unity it is impossible to distinguish between either sonic or Alfv\'enic numbers, which is to be expected.

A SNR greater than 100 is sufficient for showing differences in $w^{-2}$ and $a$ between
sub- and super-Alfv\'enic simulations, with sub-Alfv\'enic producing larger values.  
In the case of supersonic turbulence a SNR of 20 is generally sufficient to discern between 
sub- and super-Alfv\'enic gas using Tsallis fit parameters with 1.5$\sigma$ confidence
for all lags. While this is promising that all pixel scales show statistically significant 
differences, the shape of the $w^{-2}$ or $a$ curve vs. lag converges to non-noise levels only for 
lag $\sim$20 pixels and therefore, small lags may not be as reliable as larger lags in the presence of low SNR. Subsonic cases can 
distinguish Alfv\'enic number with confident in the 1 $\sigma$ range at SNR 20 and greater 
for lags larger then 32 pixels.

\subsubsection{Simulating Cloud Boundaries}
\label{cloud}
Studies of the ISM  frequently deal with both clouds-like objects as well as diffuse gas.
While our simulations are most directly applicable to diffuse
ISM, we can also study the effects the Tsallis fits will have on ISM gas with
boundaries. An example of such a situation would be molecular clouds 
which have characteristic radially decreasing density values.  To mimic this,
we convolve our 3D density simulations with a spherical function which
maintains the value inside a given radius R from the center and creates a Gaussian decay
outside this radius. For our simulations we choose an R of 205,
90, and 10 pixels corresponding to 2.5, 6, and 51 times smaller
than box size respectively. The cloud convolved 3D simulations are then
averaged along a LOS creating a column density map as described in
 section \ref{coldn}. For reference a column density map
with a radial decreasing Gaussian boundary starting at R=90 parallel to the mean magnetic field
is presented in Figure \ref{fig:3dpic} (bottom right) for simulation \#2 (${\cal M}_s$=10,
${\cal M}_A$=0.7).

\begin{figure*}[tbh]
  \begin{center}
    \includegraphics[angle=90,keepaspectratio=true,scale=0.5]
    {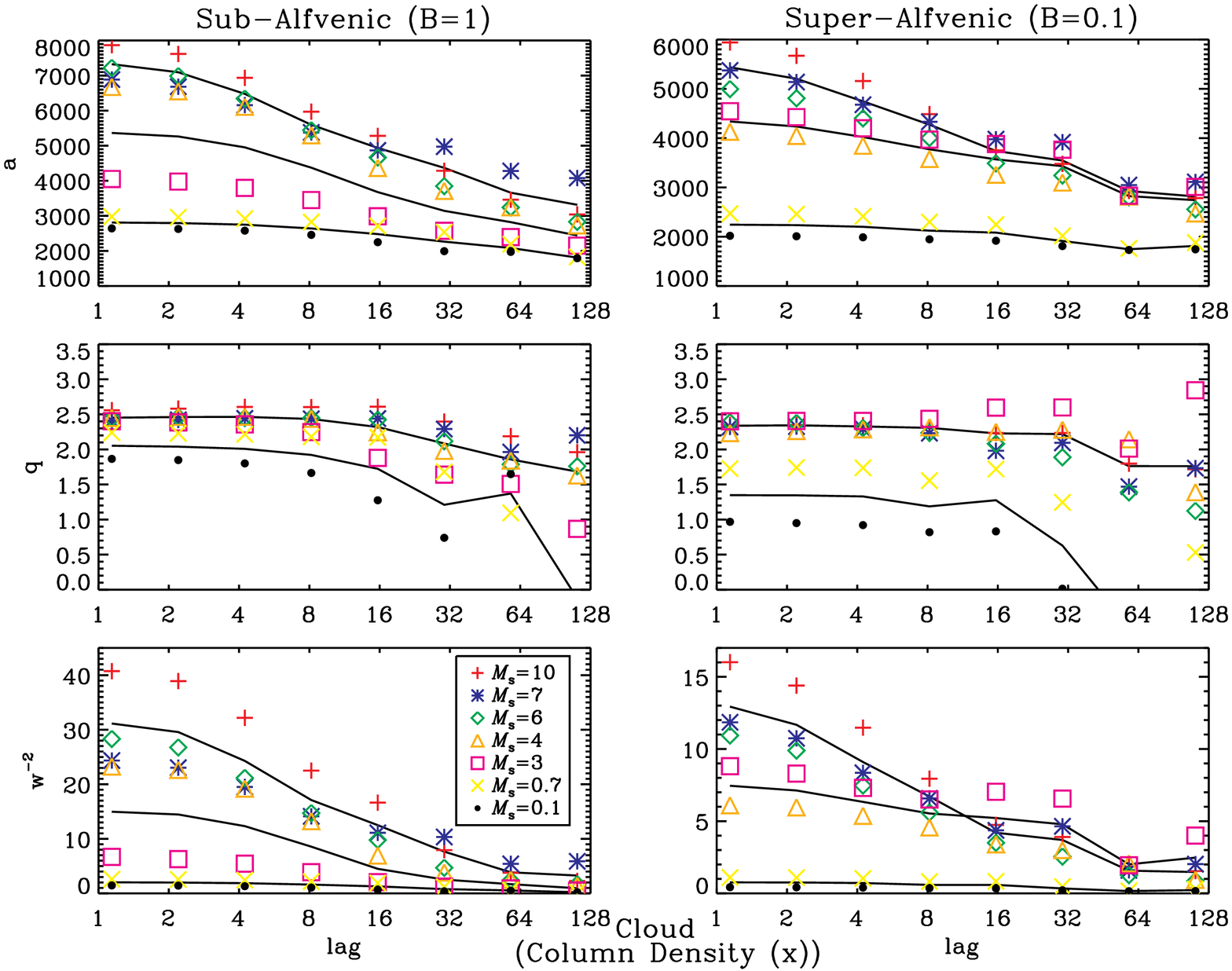}
    \caption{From top to bottom fit parameters $a$ (amplitude of fit), 
      $q$ (related to fitting tails of PDF), and $w$ (PDF width plotted 
      as $w^{-2}$) are displayed vs spatial lag for all 14 simulated cloud bounded column 
      density simulations created parallel to the mean magnetic field (R=205, 2.5 times 
      smaller than cube and the scale of the turbulent energy injection). 
      The left and right columns correspond to sub- and super-Alfv\'enic simulations
      respectively. Three solid lines are over-plotted 
      for $a$ and $w^{-2}$ averaging the supersonic (${\cal M}_s$=10, 7, 6), 
      transonic (${\cal M}_s$=4, 3), and subsonic (${\cal M}_s$=0.7, 0.1) 
      simulations. Two solid lines are over-plotted for the $q$ parameter
      averaging the supersonic (${\cal M}_s$=10, 7, 6, 4, 3) and subsonic
      (${\cal M}_s$=0.7, 0.1) simulations. Parameters $a$ and $w^{-2}$ show 
      sensitivities to ${\cal M}_s$ and ${\cal M}_A$ (note 
      scale between left and right columns). Parameter $q$ 
      is only slightly sensitive to the simulation's compressibility. Lowest error 
      are achieved for supersonic simulations at lags between 4 and 32 pixels 
      where percent errors are below 44$\%$, 40$\%$, and 84$\%$ for $a$, $q$, and $w^{-2}$ 
      respectively. Outside this constraint, error $>$ 150$\%$ for each parameter.}   
    \label{fig:cloudall}
  \end{center}
\end{figure*}

In general, a radially decreasing cloud boundary creates an increase in the
kurtosis of the PDFs as compared with unbounded column density (not shown). This change
comes from the increase in  zero point values of $\rho(x+r)-\rho(x)=0$ as more
contrast is created. At a radius of 205 pixels the PDFs are still fairly uniform and
only see slight variations from Gaussianity for subsonic simulations and fits
remain similar to those seen in  Figure
\ref{fig:cdxhists1}. 

\begin{figure*}[tbh]
  \begin{center}
    \includegraphics[angle=90,keepaspectratio=true,scale=0.5]
    {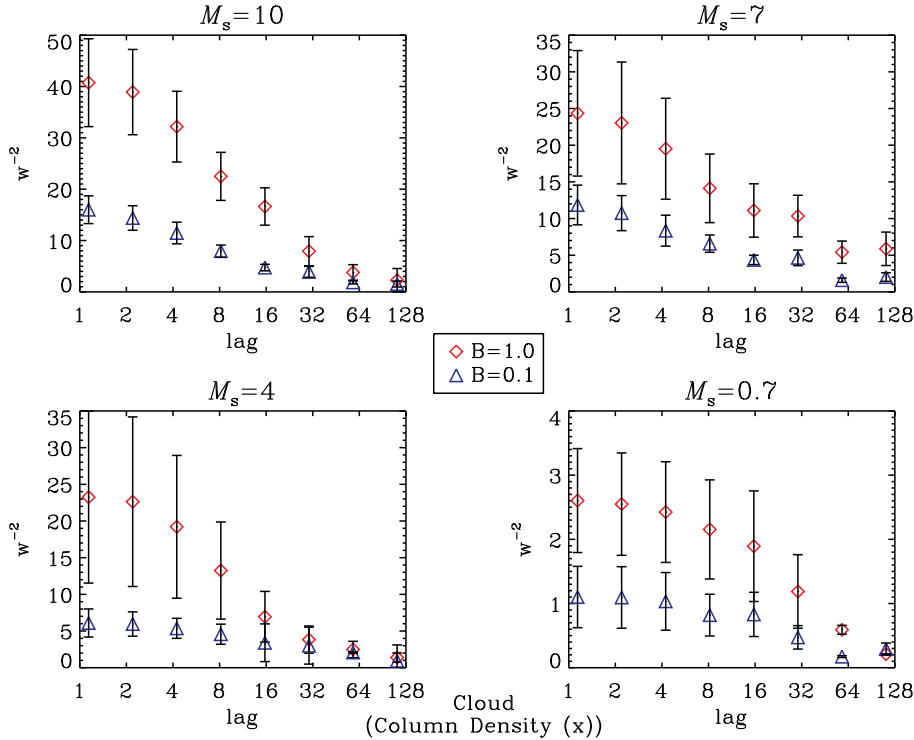}
    \caption{An enlarged version of Figure \ref{fig:cloudall}'s bottom panels 
      for ${\cal M}_s$
      = 10, 7, 4, and 0.7. Sub- and super-Alfv\'enic simulations are over-plotted in red
      diamonds and blue triangles respectively. Both Super- and Sub-sonic
      cases have elevated values for high magnetization. Error bars are calculated from the standard 
      deviation of the mean of each ${\cal  M}_s$ group (highly supersonic, mildly supersonic,
      subsonic) at each lag.}
    \label{fig:cloud4}
  \end{center}
\end{figure*}

Figure \ref{fig:cloudall}
presents the $a$ (top), $q$ (middle) and $w^{-2}$ (bottom) fit parameters for the R=205 
bounded simulations parallel to the mean magnetic field. As with 3D density and
unbounded column density, sub- and super-Alfv\'enic simulations are presented
on the left and right panels respectively. Three averaging lines are plotted
through the highly super-, mildly super-, and subsonic simulations for $a$ and $w$, while 
$q$ has only super- and subsonic averaging lines. Sensitivities to ${\cal
  M}_s$ are still apparent throughout the $a$ and $w^{-2}$ fits values and slopes but 
emphasized more in sub-Alfv\'enic simulations. ${\cal M}_A$ remains strongly tied to $w$,
as presented in Figure \ref{fig:cloud4}. In each scenario, the simulation
containing a stronger magnetic field results in a $\geq$2 times higher value
of $w^{-2}$. The $q$ parameter shows little effect from a radially
decreasing boundary condition and still is only mildly affected by
simulation compressibility.

Fit parameter errors of cloud bounded column densities shows similar trends to those seen 
in full resolution 2D density maps. Parameter $a$ maintains the lowest error from lag 2 to 
32 pixels where errors remain $<$ 16$\%$. Outside this range, errors increase and reach a 
maximum error of 46$\%$ at the 1 pixel lag. $q$ achieves errors below 36$\%$ for most 
simulation lags excluding subsonic simulations and large lags. In this regime error reaches
90$\%$. Errors of the $w^{-2}$ parameter display similar trend as $q$ where errors below 
50$\%$ are produced for mid range lags (4-32 pixels) and highly and mildly supersonic simulations. Errors 
reach $>$ 150$\%$ at small lags and subsonic simulations at high lags. This further confirms 
the trend that decreasing resolution has on the quality of fits and fit parameter values. 
Even as errors increase, the $w^{-2}$ parameter is still able to distinguish ${\cal M}_A$ within the error bars.

Similar to the incremental column density PDFs, our largest clouds proved well
described by Tsallis statistics. However, the  decrease in cloud size greatly affects
the PDFs and the ability of the Tsallis function to fit them. For example, in the case of our smallest 
cloud (R=50 pixels), the PDF and fit for simulation \#2 (${\cal
  M}_s$=10, ${\cal M}_A$=0.7) at even our largest lag (128 pixels, not shown), is extremely
kurtotic and skewed to negative values. 
This is due to radially decreasing density, the value of $\rho(x+r)-\rho(x)$ is in general $>$ 0 for
three out of the four directional calculations (+x, -x, +y, -y). Tsallis is
a symmetric  function about  $\rho(x+r)-\rho(x)=0$ and this skewness of the 
PDF prevents tight fits. While quality of the fit is no longer
intact, the relative amplitudes and widths change with lag, ${\cal M}_s$,
and ${\cal M}_A$. 

Cloud convolved column densities with LOS perpendicular to the mean
magnetic field were also analyzed and showed the same trends and no discernible
correlation with the LOS orientation.

\subsection{PPV Cubes}
\label{ppv}
The application of Tsallis statistics has proven itself to be a useful tool 
for exploring a wide parameter range
of sonic and Alfv\'enic Mach numbers in density and column density data.
However, while density fluctuations are indeed useful for characterizing
turbulence, more appropriate is the use of velocity information.  
We test the usefulness of Tsallis PDF fits on Position-Position-Velocity (PPV)
data in order to see if additional information is provided by including the
velocity axis.   We create synthetic PPV cubes of our full simulation range
using density and velocity with LOS perpendicular to mean magnetic field. In
order to test a variety of velocity resolutions we create PPV cubes with
velocity widths of 0.15 ($\sim$30 channels), 0.07 ($\sim$60 channels), and
0.007 ($\sim$600 channels).  Units remain scale free.

\begin{figure*}[tbh]
  \begin{center}
    \includegraphics[angle=90,keepaspectratio=true,scale=0.5]{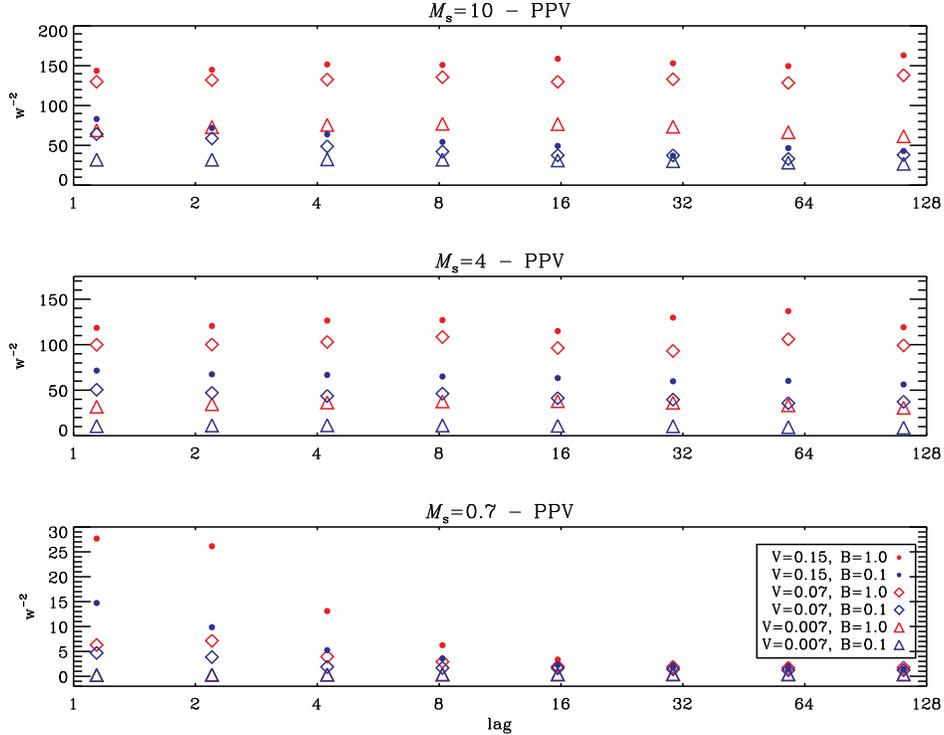}
    \caption{Parameter $w$ (PDF width plotted as $w^{-2}$) for ${\cal M}_s$= 10, 4, 0.7 from 
    top to bottom respectively. Each panel plots the sub- and super-Alfv\'enic simulations for the three
    velocity channels (see legend). A sensitivity to ${\cal M}_A$ is seen in every case but 
    more pronounces for high turbulence simulations and few velocity channels. Subsonic 
    simulations shows trends that resemble density while supersonic simulations do not.}
    \label{fig:PPV}
  \end{center}
\end{figure*}

Figure \ref{fig:PPV} presents
$w^{-2}$ vs. lag for the sub- and super-Alfv\'enic versions of the ${\cal M}_s$=10, 4, and 0.7
simulations from top to bottom respectively. Each panel has the three velocity resolutions
over-plotted for comparison. Low, middle, and high resolutions are represented by the 
circle, diamond, and triangle respectively. Within each resolution, red symbols correspond 
to sub-Alfv\'enic ($B_\mathrm{ext}$=1.0) while blue corresponds to super-Alfv\'enic ($B_\mathrm{ext}$=0.1). 
All three panels display a common trend of
decreasing $w^{-2}$ value as the velocity resolution increases
(width of the velocity channels decreases). Also, it is clear that the $w$
parameter is sensitive to the ${\cal M}_A$ number for PPV simulations and produces
elevated values for simulations with high magnetization. 

The general shape of $w^{-2}$ over increasing lags is distinctly different for 
PPV data than in density, velocity, or magnetic field. The PDFs, and therefore fit parameters
are more consistent with lag variations and produce smoother trends. This alternative trend 
is only seen in supersonic PPV data while values of $w^{-2}$ for subsonic PPV data more closely
resemble the trends seen in density analysis.

Parameter $a$ shows minimal variations to ${\cal M}_A$ or ${\cal M}_s$ for PPV simulations. 
Sensitivities to ${\cal M}_s$ are seen on a scale too small to be used observationally and 
there is no coherent relationship to ${\cal M}_A$. Similar to previous discussions $q$ only 
provides limited information on compressibility and none on ${\cal M}_A$.

It should be noted that during the distribution creation process, especially
for high velocity resolutions, velocity channels with little emission were
excluded. In addition, a majority of the small lag PDFs had a single multiple order of magnitude jump
at the center of the distribution (where $\rho(x+r)-\rho(x)=0$). This point was
excluded from the fitting process. This is simply due to the way the emission is spread over the channels in PPV with a Maxwellian velocity distribution.
Most gas is near the mean velocity and a Gaussian decrease in emission is seen as one goes to the extreme minimum and maximum channel
velocities.  

Analysis of density distributions along the velocity axis for individual pixels were also 
analyzed. An attempt to fit these distributions with the Tsallis function proved ineffective. 
We find the distribution of density emission along velocity channels, especially for high 
turbulence, are highly skewed and irregular which is consistent with the findings of Falgarone 
et al. (1994). As Tsallis is a symmetric function it is not able to provide tight fits to these
highly irregular and varying distributions. 

\section{DISCUSSION}
\label{disc}
\subsection{Summary and the Relation of Tsallis to Other ISM Statistics}
We studied Tsallis statistics as a way to characterize MHD ISM turbulence. While we do not address 
the theoretical justification of the particular form of the fits predicted by the Equation (1), 
our study shows that this fit provides a very good correspondence with experimentally obtained
distributions. We showed that the parameters that enter the Tsallis expression are dependent on
the Sonic Mach (${\cal M}_s$) and Alfv\'enic Mach (${\cal M}_A$) numbers, confirming and 
extending the earlier claims of Esquivel \& Lazarian (2010). In addition, we explored the 
differences in the dependencies of the fitting parameters on  both ${\cal M}_A$, 
${\cal M}_s$ and on the resolution. This allows one to find ${\cal M}_s$ and ${\cal M}_A$ 
independently. We also applied Tsallis statistics to PPV data, which, as far as we know, 
is the first time such a study has been undertaken.

Another important point that we address here is the influence of telescope resolution in the form of smoothing. 
This issue was largely omitted in our earlier publications, but our results show that 
the influence of the telescope beam  may be important for observational studies. Not only does smoothing affect
the statistics, but it could actually be a helpful addition to the statistical procedure to study Mach numbers.
Indeed, telescope measurements introduce their own smoothing, 
which depends on the telescope resolution. As the injection scale of the turbulence is 
frequently not known, the given measurements of the parameters of the Tsallis 
statistics may not be straightforward to identify with the particular simulation. In 
this situation, additional smoothing can help to see to what parameters of turbulence 
the observed data corresponds to.

Burkhart et al. (2010) studied the Small Magellanic Cloud (SMC) to characterize the 
properties of turbulence in its individual parts. The same approach is applicable with 
Tsallis statistics. It is convenient to note, that one can gain information on regions 
turbulence or magnetization without knowing the precise resolution of the telescope. Our
motivation for applying boundaries to our simulations was fueled by making Tsallis applicable to 
more compact objects, such as those that may be found in molecular clouds. 
Our results show clear tendencies to determine Mach number information from the Tsallis 
fitting parameters.

In addition, our study shows that the use of the full PPV data can bring additional 
advantages. For instance, the evolution of Tsallis parameters with  is clearly different for 
subsonic and supersonic cases (see Figure \ref{fig:PPV}) in PPV data.

We feel that the Tsallis fit provides useful way of studying turbulence, which should 
be implemented along with other techniques used previously (see Kowal, Lazarian, \& Beresnyak 2007; 
Burkhart et al. 2009, 2010). We advocate an approach which combines the use of different 
techniques. For instance, the effect of the telescope beam smoothing depends on the ratio of 
the angular size of the turbulence energy scale in the object under study and the telescope 
resolution. The former can be obtained by studying the power spectra of turbulence.

While we are still searching for the basic approaches to study turbulence observationally, 
the successes in theoretical and numerical studies provide us with a reliable guidance for 
obtaining statistics of turbulence from observations. The development of the VCA
and VCS techniques (see Lazarian \& Pogosyan 2000, 2004, 2006, 2008, Chepurnov \& Lazarian 
2008) for instance, quantified how velocity and density create structures observed in the PPV. 
Results on the anisotropies of observational data statistics in the plane of
the sky in  Lazarian, Pogosyan \& Esquivel (2002) and subsequent publications 
(Esquivel \& Lazarian 2005) reveals strong dependencies on media magnetization. 
Kurtosis and skewness measures of the density (Kowal, Lazarian, \& Beresnyak 2007; Burkhart et al. 2009) 
provide good insight mostly into the intensity of turbulence, i.e. its  ${\cal M}_s$. Genus 
(see Lazarian 1999, Kim \& Park 2007, Chepurnov et al. 2008) provides a measure of topology, 
while bispectrum (see Burkhart et al. 2009) provides insight into mode correlation and phase 
information of the turbulence\footnote{The latter two measures were borrowed from 
cosmological studies and their use for ISM studies was proposed in Lazarian (1999). 
The process of appreciating of their value for the ISM is currently under way.}. 
Additional information can be obtained by employing other tools, e.g. dendrograms 
(Rosolowsky et al. 2008, Goodman et al. 2009, Burkhart et al. 2011). At the same time the 
synergy of the simultaneous use of different techniques is still to be explored.

\subsection{Relation of This Study to Previous Tsallis Work}
\label{relate}
In this paper we fit the Tsallis distribution function to PDFs of turbulence with 
little or no theoretical justification.  Indeed, while 
the history of turbulence studies can be traced back to Kolmogorov (1941), turbulence studies have
since taken many different directions.  In this paper, we are particularly interested 
in characterizing magnetized turbulence in the ISM.   The Tsallis distribution 
formalization comes from a different background, specifically the issue of quantities 
which are invariant in the Navier-Stokes equations at high Reynolds numbers and the 
assumption that the singularities due to the invariance distribute themselves multifractally
in physical space.  Much work has been done to address both the theoretical frame work and 
application of the Tsallis statistics to numerical and laboratory turbulence PDFs
(see Tsallis 1999 and Arimitsu \& Arimitsu (2000a, 2000b, 2001, 2002, 2003)).  For example, in Arimitus \&  
Arimitus 2003, the authors successfully fit the Tsallis distribution to PDFs of turbulence
in three different laboratory experiments.  Burlaga and collaborators successfully fit 
Tsallis to solar wind data (Burlaga, Vi{\~n}as, \& Wang 2009). 

In this paper, we successfully fit Tsallis 
to our numerical magnetized ISM turbulence, even in the case of observable quantities such 
as column density and PPV emission cubes.   In the case of the laboratory, space observations, 
and numerical experiments; the geometry of the system, the Reynolds numbers, and the type of 
turbulence in question, are all very different.  Never-the-less, in all cases the Tsallis 
distribution provided good fits to the data.  Motivated by these past studies, we applied 
the  Tsallis function in an \emph{empirical} way to determine the parameters of turbulence, and 
we direct the curious reader to the literature mentioned above for a more in depth treatment 
of the theoretical foundations of the Tsallis function.  Also, our paper is motivated towards 
making Tsallis statistics useful for observational studies of turbulence, and in this case 
the discussion of multifractal theory is beyond our current scope. We do not attempt to justify 
or even test these ideas, but investigate to what degree the fit provided by the two parameter 
Tsallis formula corresponds to our numerical data.

\subsection{The Issues of Numerical Resolution}
\label{numissue}
While there is no doubt that substantial progress has been made in last decade in numerical 
simulations of turbulence, the community still has a long way to go in terms of achieving the 
resolution required to fully capture ISM physics. One issue in particular is the discrepancy 
between astrophysical values of Reynolds (Re) and magnetic Reynolds (Rm) and the numerical values. 
As a rule, astrophysical environments are turbulent and astrophysical turbulence is 
characterized by enormously large Re and Rm.  For instance, the Rm number, which characterizes 
the degree of frozenness of magnetic field within eddies, may differ in astrophysical environments 
and numerical simulations by a factor larger than $10^{10}$.  With these sorts of discrepancies 
in mind (which exist for all numerical codes that simulate turbulence), how are we able to 
justify our results?

There are several avenues to justify the independence of our results from Re and Rm values. 
The particular numerical code used here provides enough resolution to allow us to see the 
inertial range of turbulence. If turbulence is evolving along the inertial range it is 
self-similar, and the change in Re and Rm does not change the structure of the large scale 
motions which we study here. Indeed, according to the Lazarian \& Vishniac (1999) model of turbulent 
reconnection (tested in Kowal et al. 2009), magnetic reconnection in 
a turbulent media depends on the large scale field wondering determined by large scale motions. 
Thus we do not expect the unresolved small scales to affect the physics of our simulations at 
large scales that we study. Keeping all of this in mind, we also tested our Tsallis fits at 
resolutions of $256^3$ and $512^3$ and found no deviation in how well Tsallis was able to both 
fit the PDFs and describe the turbulence in question.

\section{CONCLUSIONS}
\label{conc}
In this paper we applied the Tsallis formalism to
simulated diffuse ISM isothermal ideal MHD turbulence using incremental PDFs of density,
velocity, and magnetic field. This method was also applied to simulated column
densities with measures taken to duplicate observed measurements such as
smoothing, noise, and radial decreasing cloud boundaries. The use of Tsallis
statistics on PPV data was also explored. A summary of our findings is as
follows: 

\begin{itemize}
\item The Tsallis function is capable of well describing incremental PDFs of
  a wide range of simulated MHD density, velocity, and magnetic field with
  its three fit parameters $a$ (amplitude), $w$ (width), and $q$ (related to the
  PDF's tails).

\item For 3D density, fit parameters $a$, $q$, and $w^{-2}$ are sensitive to
  ${\cal M}_s$. $a$ and $w^{-2}$ show sensitivities to ${\cal M}_A$ with
  both displaying  greater values for sub-Alfv\'enic simulations. 

\item Magnetic field and velocity show sensitivities to ${\cal M}_s$
  and  ${\cal M}_A$ but these are not as strong as density.

\item PDFs of column density are equally well described by the Tsallis distribution. 
  Fit parameters $a$, $q$, and $w^{-2}$ remain sensitive to ${\cal M}_s$. $a$
  looses some of its magnetic sensitivity but $w^{-2}$ consistently produces
  elevated values for sub-Alfv\'enic simulations.

\item The affect of degrading resolution lowers both 
  $a$ and $w^{-2}$ vs. lag monotonically. Supersonic simulations are most
  affected by smoothing. $a$ and $w^{-2}$ remain sensitive to ${\cal M}_s$ and
  ${\cal M}_A$ for mild smoothing. $q$ looses ${\cal M}_s$ sensitivity with
  smoothing. 

\item The addition of smoothing at varying degrees acts as an additional method to 
  constrain sonic Mach number through the analysis of the slope of 
  parameters $a$ or $w^{-2}$ over lag.

\item Noise has the largest affect on fit parameters. Small scale variations
  strongly alter the PDFs of low lags and the shape of the fit
  distributions. Sensitivities to ${\cal M}_A$ and ${\cal M}_s$ are still
  seen on smaller scales as the spatial lag increases.

\item Radially deceasing cloud boundaries have little affect as long as a
  there are enough points inside providing good signal. As clouds become
  smaller, PDFs become more kurtotic and skewed preventing meaningful fits.
  
\item PPV data was fit with Tsallis distribution excluding the zero point and 
parameter $w^{-2}$ showed sensitivities to
  ${\cal M}_s$ and ${\cal M}_A$.

\item Tsallis statistics of incremental PDFs is a successful tool in describing a  
  wide range of high resolution MHD simulations and their observational counter parts. 
  Along with its sensitivities to turbulence and magnetic fields, the Tsallis 
  distribution is highly complimentary to power spectrum and other ISM statistical tools.
\end{itemize}

\bibliographystyle{apj}

\acknowledgments
Authors wish to thank A. Esquivel for use of the Tsallis code and productive discussion and 
the anonymous referee for thoughtful comments.
B.T. is supported by the NSF funded Research Experience for Undergraduates
(REU) program through NSF award AST-1004881. B.B. acknowledges the
NSF Graduate Research Fellowship and the NASA Wisconsin Space Grant
Institution.  A.L. thanks both NSF AST 0808118 and the Center for Magnetic
Self-Organization in Astrophysical and Laboratory Plasmas.  

\end{document}